\documentclass[sigplan,screen]{acmart}

\acmYear{2024}\copyrightyear{2024}
\setcopyright{acmlicensed}
\acmConference[PPoPP '24]{The 29th ACM SIGPLAN Annual Symposium on Principles and Practice of Parallel Programming}{March 2--6, 2024}{Edinburgh, United Kingdom}
\acmBooktitle{The 29th ACM SIGPLAN Annual Symposium on Principles and Practice of Parallel Programming (PPoPP '24), March 2--6, 2024, Edinburgh, United Kingdom}
\acmDOI{10.1145/3627535.3638474}
\acmISBN{979-8-4007-0435-2/24/03}

\usepackage{booktabs}   
\usepackage{subcaption} 

\usepackage{soul}

\usepackage{xcolor}

\usepackage{amsmath,amsfonts}
\usepackage{nccmath}

\definecolor{olivegreen}{rgb}{0, 0.6, 0}
\definecolor{redorange}{HTML}{FF5349}
\definecolor{blue(ncs)}{rgb}{0.0, 0.53, 0.74}
\definecolor{navy}{HTML}{273BE2}
\definecolor{darkpastelpurple}{rgb}{0.59, 0.44, 0.84}
\definecolor{darkolivegreen}{rgb}{0.33, 0.42, 0.18}
\definecolor{darkgreen}{rgb}{0.0, 0.2, 0.13}
\definecolor{darkpastelgreen}{rgb}{0.01, 0.75, 0.24}

\definecolor{black}{HTML}{000000}
\definecolor{white}{HTML}{ffffff}
\definecolor{color1}{HTML}{ACE5EE}
\definecolor{color2}{HTML}{0093AF}
\definecolor{color3}{HTML}{CC0000}
\definecolor{color4}{HTML}{0087BD}
\definecolor{color5}{HTML}{333399}
\definecolor{color6}{HTML}{20B2AA}

\usepackage{enumitem}
\setlist[enumerate]{leftmargin=5mm} 
\setlist[itemize]{leftmargin=5mm}

\usepackage{xspace}

\usepackage{graphicx}

\usepackage[skip=5pt]{caption} 

\usepackage[export]{adjustbox}
\usepackage{wrapfig}
\usepackage{tikz}

\usepackage{verbatim}

\usepackage{multicol}
\usepackage{multirow}
\usepackage{makecell}
\usepackage{bbm}

\usepackage{algorithm}
\usepackage{algpseudocode}
\algnewcommand{\IfThen}[2]{
  \State \algorithmicif\ #1\ \algorithmicthen\ #2\ }
\algnewcommand{\IfThenElse}[3]{
  \State \algorithmicif\ #1\ \algorithmicthen\ #2\ \\ \algorithmicelse\ #3\ }

\usepackage[noabbrev, capitalize]{cleveref} 
\Crefname{section}{\S}{\S}
\usepackage{hyperref} 

\usepackage{pifont}

\usepackage{ulem}

\DeclareMathOperator*{\argmax}{argmax} 

\newcommand{\x}{\texttimes{}\xspace}

\newcommand{\thiswork}{AGAThA\xspace}
\newcommand{\namewhy}{\thiswork stands for ``A GPU Acceleration for Third-generation sequence Alignment''. The four types of DNA bases (A, G, C, and T) are capitalized.}
\newcommand{\bandwidth}{We use the term ``band width'' to represent the width of the diagonal band in the score table, which is different from ``memory bandwidth''.}

\newcommand{\SD}{Sliced Diagonal\xspace}
\newcommand{\Sd}{Sliced diagonal\xspace}
\newcommand{\sd}{sliced diagonal\xspace}

\newcommand{\SR}{Subwarp Rejoining\xspace}
\newcommand{\Sr}{Subwarp rejoining\xspace}
\newcommand{\sr}{subwarp rejoining\xspace}

\newcommand{\UB}{Uneven Bucketing\xspace}
\newcommand{\Ub}{Uneven bucketing\xspace}
\newcommand{\ub}{uneven bucketing\xspace}

\newcommand{\MT}{Rolling Window\xspace}
\newcommand{\Mt}{Rolling window\xspace}
\newcommand{\mt}{rolling window\xspace}

\newcommand{\WS}{Window Spilling\xspace}

\newcommand{\rev}[1]{{\color{black}#1}}

\setstcolor{olivegreen}

\newcommand{\select}[2]{#2}

\definecolor{carminered}{rgb}{1.0, 0.0, 0.22}
\definecolor{neonpink}{RGB}{255, 138, 216}
\definecolor{bandyellow}{RGB}{255, 175, 0}

\newcommand{\midcircled}[1]{\raisebox{.5pt}{\textcircled{\raisebox{-.9pt} {#1}}}}

\begin{document}

\normalem

\title[\thiswork]{\thiswork: Fast and Efficient GPU Acceleration of Guided Sequence Alignment for Long Read Mapping}         


\settopmatter{authorsperrow=3}
\author{Seongyeon Park}
\affiliation{%
  \institution{Seoul National University}
  \city{Seoul}
  \country{South Korea}
}
\email{syeonp@snu.ac.kr}
\orcid{0009-0007-3480-6626}

\author{Junguk Hong}
\affiliation{%
  \institution{Seoul National University}
  \city{Seoul}
  \country{South Korea}
}
\email{junguk16@snu.ac.kr}
\orcid{0009-0001-4004-7714}

\author{Jaeyong Song}
\affiliation{%
  \institution{Seoul National University}
  \city{Seoul}
  \country{South Korea}
}
\email{jaeyong.song@snu.ac.kr}
\orcid{0000-0001-9976-7487}

\author{Hajin Kim}
\affiliation{%
  \institution{Yonsei University}
  \city{Seoul}
  \country{South Korea}
}
\email{kimhajin@yonsei.ac.kr}
\orcid{0009-0004-3709-1572}

\author{Youngsok Kim}
\affiliation{%
  \institution{Yonsei University}
  \city{Seoul}
  \country{South Korea}
}
\email{youngsok@yonsei.ac.kr}
\orcid{0000-0002-1015-9969}

\author{Jinho Lee}
\affiliation{%
  \institution{Seoul National University}
  \city{Seoul}
  \country{South Korea}
}
\email{leejinho@snu.ac.kr}
\authornote{Corresponding author.}
\orcid{0000-0003-4010-6611}

\begin{abstract}
With the advance in genome sequencing technology, the lengths of deoxyribonucleic acid (DNA) sequencing results are rapidly increasing at lower prices than ever. 
However, the longer lengths come at the cost of a heavy computational burden on aligning them.
For example, aligning sequences to a human reference genome can take tens or even hundreds of hours.
The current de facto standard approach for alignment is based on the guided dynamic programming method.
Although this takes a long time and could potentially benefit from high-throughput graphic processing units (GPUs), the existing GPU-accelerated approaches often compromise the algorithm's structure, due to the GPU-unfriendly nature of the computational pattern.
Unfortunately, such compromise in the algorithm is not tolerable in the field, because sequence alignment is a part of complicated bioinformatics analysis pipelines.
In such circumstances, we propose AGAThA, an exact and efficient GPU-based acceleration of guided sequence alignment.
We diagnose and address the problems of the algorithm being unfriendly to GPUs, which comprises strided/redundant memory accesses and workload imbalances that are difficult to predict.
\rev{According to the experiments on modern GPUs, AGAThA achieves 18.8$\times$ speedup against the CPU-based baseline, 9.6$\times$ against the best GPU-based baseline, and 3.6$\times$ against GPU-based algorithms with different heuristics.}
\end{abstract}



\maketitle


\section{Introduction}

\label{sec:intro}

Genome sequence analysis has largely impacted our lives, from aiding medical fields
to others such as epidemiology~\cite{parkhill2011bacterial}, agriculture~\cite{bohra2020genomic}, and even basic science~\cite{suzuki2020advent}.
Such analysis is made possible by sequencing, which generates sequences called \emph{reads} from strands of deoxyribonucleic acid (DNA) extracted from specimens. 
Although there has been a wide variety of sequencing techniques, the recently introduced third-generation sequencing (TGS) technique produces very long reads compared to its predecessors. 
While previous generation sequencing generated short reads of around 150$\sim$300 bps (base pairs), TGS can produce reads longer than an average of 10 kbps and higher~\cite{giani2020long, mantere2019long}. 
By generating longer and high-quality reads, TGS sheds light on new attributes and genomic mutations that were difficult to spot before.

However, this drastic increase in size comes at the cost of sharply increased time spent on the sequence analysis.
The major essential step of the sequence analysis is read alignment, which aligns reads (i.e., \emph{queries}) 
to reference genomes to find the location of each read within the whole DNA~\cite{alser2021technology}. 
Due to the larger data size,
prior read alignment algorithms struggle to process reads with acceptable low latency.

There are two de facto standard methods for aligning reads: Minimap2~\cite{minimap2} for longer reads from TGS, and BWA-MEM~\cite{bwamem} for shorter reads from the previous generation.
At their cores, they form a very similar structure, based on a \emph{guided dynamic programming} approach.
Despite being highly optimized, \select{Minimap2 has a}{they exhibit} very long execution time even with modern multi-core CPUs. 
For example, \cite{alser2022molecules} shows that mapping TGS reads to the entire human genome (3.2 GB) \select{}{using Minimap2} can take 49 hours. 
To amend this, there have been several attempts to accelerate it on powerful GPUs. 

\rev{However, to the extent of our knowledge, existing attempts compromise the structure~\cite{manymap} or target different alignment algorithms~\cite{saloba, gasal2, zeni2020logan}.} 
This is problematic because read alignment is not a stand-alone application, but an early part of complicated bioinformatics pipelines. 
Any change in this step would require an immense amount of verification on the entire pipeline.
Therefore, the exactness of the algorithm cannot be traded off for speedup. 
\rev{This indicates that there is an urgent need for accelerating guided dynamic programming on modern GPU architectures.}

\rev{We reveal that the difficulty in implementing the guided programming algorithm lies in the large number of random memory accesses and the severe dynamic workload imbalance caused by the guided alignment algorithm.}
Based on this diagnosis, we propose \thiswork\footnote{\namewhy}, a GPU acceleration method for an exact and fast implementation of the guided alignment algorithm. 
To the best of our knowledge, our method is the first to accelerate the exact reference algorithm, achieving both exactness and speed.  
First, we propose an efficient scheme to calculate the termination condition of the guided alignment. 
The proposed scheme transforms the memory access patterns to be more sequential and coalesce better on the GPU memory hierarchy. 
Second, we propose a tiling scheme for the solution space to reduce the number of redundant memory accesses and unnecessary computations, 
which both depend on the execution order. 
Third, we design a method for mitigating the workload imbalance that is difficult to predict at both intra- and inter-warp levels. 
At the intra-warp level, we utilize a form of work stealing that is tightly coupled with the sequence alignment algorithm.
At the inter-warp level, we observe a long-tail-like distribution of the workloads and devise a way to group them such that each group contains a similar amount of workload.

Our contributions can be summarized as follows: 
\begin{itemize}
    \item {We propose \thiswork, the first exact GPU acceleration of the reference guided alignment. \rev{\thiswork not only exactly accelerates said algorithm but also significantly outperforms existing methods.}}
    \item We devise two schemes, \mt and \sd, to efficiently calculate termination conditions.
    \item We devise \sr and \ub to address the intra-warp and inter-warp load imbalance. 
    \item{\rev{We evaluate AGAThA against numerous baselines to show 9.6$\times$ speedup against the best GPU baseline implementations that target guided dynamic programming, and 3.6$\times$ against other GPU-based heuristics.}}
\end{itemize}


\section{Preliminary}
\subsection{Sequence Alignment}
\label{sec:sw}

\begin{figure}
    \centering
    \includegraphics[width=\columnwidth]{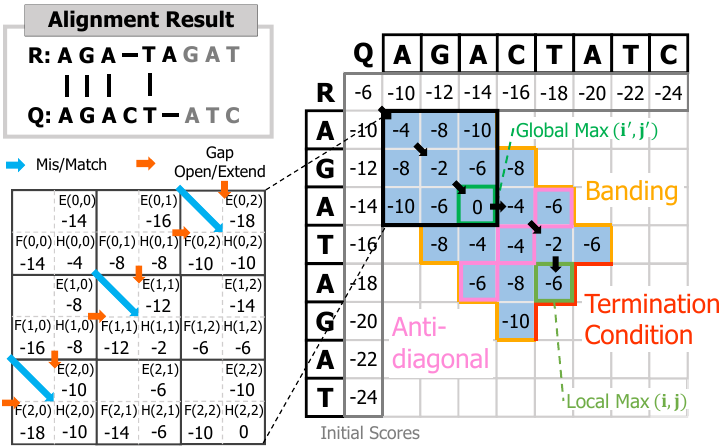}
    \caption{Guided sequence alignment.} \vspace{-3mm}
    \label{fig:sw}
\end{figure}

\textbf{Problem Definition.}
The sequence alignment problem is a variant of approximate string matching. 
Given a pair of \emph{reference} and \emph{query} strings composed of five literals `A', `G', `C', `T', and `N', it has to score how similar the two inputs are.
Unlike exact string matching problems where the difference of a single character means that the pairs do not match, 
approximate string matching has to track more possibilities of insertion (query string has an extra character compared to the reference), deletion (query string has one less character), or a simple mismatch as depicted in the top left of \cref{fig:sw}.
Additionally, a gap can be defined as one or more continuous insertions/deletions.
A gap has to be first initiated at a certain cell with an insertion/deletion (`gap open') and can be extended by adjacent insertions/deletions (`gap extend').
As a result, it outputs the alignment score, which represents the magnitude of similarity between the two sequences.

\noindent\textbf{Dynamic Programming Approach.}
The sequence alignment problem is often handled with dynamic programming~\cite{sw, nw}. 
As shown in \cref{fig:sw}, it involves filling a two-dimensional score table (right) from the reference (\textbf{R}) and query (\textbf{Q}), whose computational and space complexity is $O(N^2)$. 
Each cell in position $(i,j)$ represents $H(i,j)$, the best score that can be obtained by evaluating all possibilities until the $i$-th character of the reference string and the $j$-th character of the query string. 
$H(i,j)$ is recursively defined as:
{\small
\begin{align}
    H(i,j) &= max\{
    E(i,j),
    F(i,j),
    H(i-1,j-1) + S(R[i],Q[j])
    \}    ,\label{eq:sm}\\
    E(i,j) &= max\{
    H(i-1,j) - \alpha,
    E(i-1,j)-\beta
    \},\label{eq:sm-e}\\
    F(i,j) &= max\{
    H(i,j-1) - \alpha,
    F(i,j-1)-\beta
    \}. \label{eq:sm-f} 
\end{align}}

$S(R[i],Q[j])$ compares the $i$-th reference character and the $j$-th query character and returns positive on a match (e.g., $+2$) and negative on a mismatch (e.g., $-4$).
$E$ and $F$ are additional scores kept for tracking gaps, each storing deletions and insertions. 
$\alpha$ (e.g., $4$) and $\beta$ (e.g., $2$) are gap opening and gap extending penalties, respectively. 

From the above equations, we can find that calculating a score in the cell has dependencies on the cells from the top ($(i-1,j)$, \cref{eq:sm-e}), left ($(i,j-1)$ \cref{eq:sm-f}), and top-left ($(i-1,j-1)$, \cref{eq:sm}) values in the table, as shown in \cref{fig:sw} (bottom-left).
These necessary values are referred to as \emph{intermediate values}, and lead to the popular anti-diagonal parallelism.
By defining anti-diagonal cell group ({\color{neonpink}pink cells}) as a group of cells that have the same index sum (e.g., $H(i-1,j)$ and $H(i,j-1)$ are on anti-diagonal $(i+j-1)$), there is no dependency between calculating the score of cells in the same anti-diagonal. 
Finally, the black arrows between cells represent the locations of mis/matches and gaps to reach the maximum score in the last anti-diagonal, which is equivalent to the alignment result on the top-left of \cref{fig:sw}.

\noindent\textbf{Guiding Strategy.}
For long input reads, the aforementioned alignment algorithm is known to return many false-positive matches. 
It might find a streak of matches with large offsets (e.g., the beginning of reference against the end of the query), or after too many mismatches. 
Those are often reported to be false positives even when they output high scores~\cite{minimap2}. 
Because of this, the guiding strategy aims to filter false positive matches based on heuristics as shown in \cref{fig:sw}.

\begin{figure}
    \centering
    \includegraphics[width=.95\columnwidth]{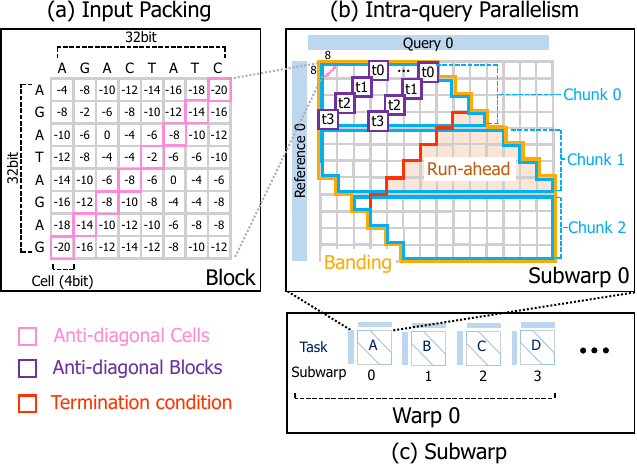}
    \caption{Existing GPU-accelerated sequence alignments.}
    \label{fig:saloba} \vspace{-3mm}
\end{figure}

\emph{Banding} ({\color{bandyellow}yellow}) is based on the idea that if there are too many insertions or deletions, 
it would be a false positive. 
For $k$-banding~\cite{bwamem, minimap2, smsmap, fujiki2020seedex}, only a diagonal band with band width\footnote{\bandwidth} $k$ ($3$ in this example) is calculated and the rest are disregarded.
In addition, \emph{termination condition} ({\color{red}red}) 
limits too many mismatches.
If the difference between the global maximum score ({\color{olivegreen}dark green}) and the current score ({\color{darkpastelgreen}light green}) becomes larger than a threshold, the termination condition is met.  
\rev{With some variants, a widely adopted form of termination condition~\cite{minimap2, minimap2avx, sadasivan2023accelerating} is}:
{\small
  \setlength{\abovedisplayskip}{3pt}
  \setlength{\belowdisplayskip}{\abovedisplayskip}
  \setlength{\abovedisplayshortskip}{0pt}
  \setlength{\belowdisplayshortskip}{2pt}
    \begin{align}
        \exists c < \mathit{len(query)} + \mathit{len(ref)}&, \quad\text{which satisfies} \\
    \mathit{ i' < i, j' < j, H(i',j') - H(i,j)} &\mathit{> Z + \beta\cdot|(i-i')-(j-j')|,} \label{eq:drop} 
    \end{align}    \begin{align}
    \text{where}\qquad (i,j) &= \argmax_{i+j=c}H(i,j), &\text{   (Local max.)}\label{eq:localmax} \\
    (i',j') &= \argmax_{i'+j'<c}H(i',j'). &\text{   (Global max.)}\label{eq:globalmax} 
    \end{align}}

In \cref{eq:drop}, $Z$ is an algorithm-specific and user-defined threshold, and $\beta$ is the gap extension score from \cref{eq:sm}.
For CPUs, this technique not only reduces false positives but also increases throughput despite the overhead of checking conditions, as it can stop calculation accordingly.
However, this causes a huge performance overhead on GPUs (see \cref{sec:moti:diagnosis}), likely explaining the absence of an exact implementation.

\subsection{State-of-the-art GPU Acceleration}
\label{sec:back:saloba}

In this section, we explain the techniques used in the state-of-the-art methods~\cite{adept, gasal2, saloba} for accelerating sequence alignment on GPUs with CUDA support.

\noindent\textbf{Input Packing.} 
Input packing~\cite{gasal2} can help deal with 
memory bandwidth bottlenecks in GPU kernels.
Because there are only 
five 
literals in genome sequences, four bits suffice for encoding each literal. 
Since GPUs typically use 32-bit words, sequences are packed with 8 literals per word.
To efficiently process this, the score table is configured in units of \textbf{block}s comprising 8$\times$8 \textbf{cell}s, which forms the smallest unit for workload distribution as depicted in \cref{fig:saloba} (a).

\noindent\textbf{Intra-query Parallelism.} 
Intra-query parallelism~\cite{saloba, adept} is an essential technique to exploit massive parallelism on the GPU. 
It assigns multiple threads to an alignment task as shown in \cref{fig:saloba} (b). 
The four threads ({\color{darkpastelpurple}purple}) 
compute four blocks concurrently, utilizing anti-diagonal parallelism (\cref{sec:sw}). 
Upon completing each block, the threads move one block horizontally in sync until it reaches the end of the band (rows 0 to 3). 
This set of blocks processed in one horizontal pass is called a \emph{chunk}.
Then, the threads move on to the next chunk (rows 4 to 7) and compute horizontally until they fill the necessary score table cells.

\noindent\textbf{Subwarp.}
In GPU kernels, it is common to exploit parallelism in units of a warp (i.e., 32 GPU threads).
However, this yields a large external fragmentation at the start and end of a chunk. 
Since threads calculate each block on the same anti-diagonal in sync,
high-numbered threads start late and low-numbered threads end early.
This can be mitigated by splitting a warp into smaller subwarps~\cite{vwc, saloba} and assigning a task to each subwarp as in \cref{fig:saloba} (c). 
This reduces external fragmentation at the cost of warp divergence (internal fragmentation), but the benefits surpass the penalties.


\section{Motivation}
\label{sec:moti}

\begin{figure*}[t]
    \centering
    \subfloat[] {
        \includegraphics[width=\columnwidth]{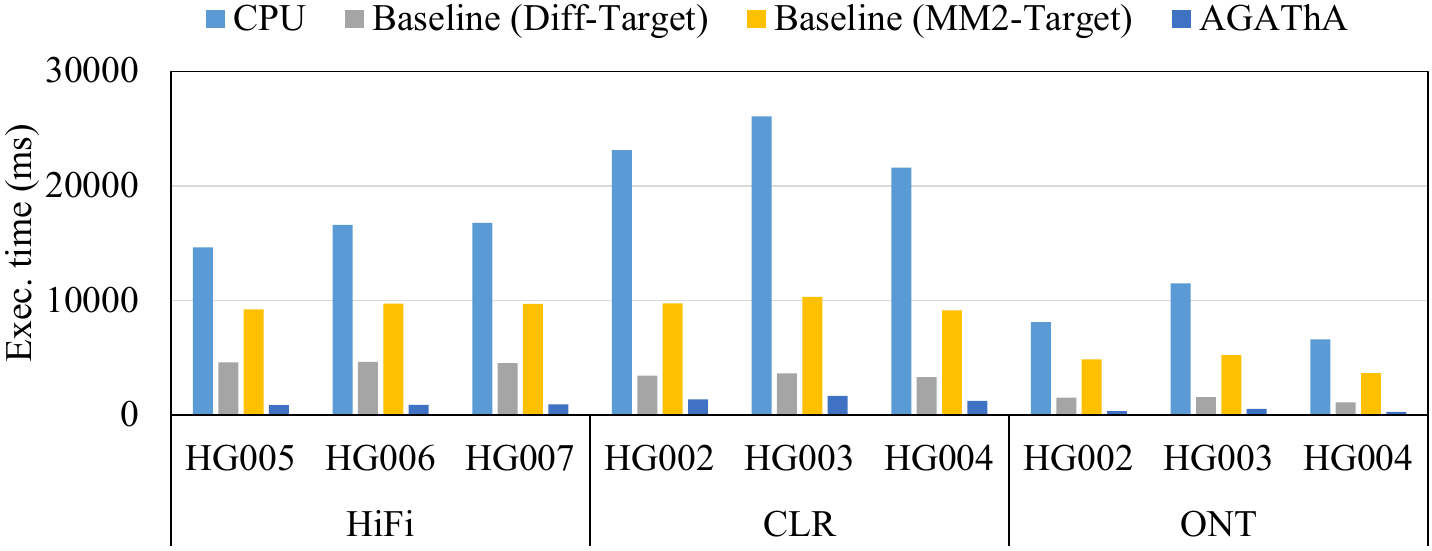}
    }
    \subfloat[]{
        \includegraphics[width=\columnwidth]{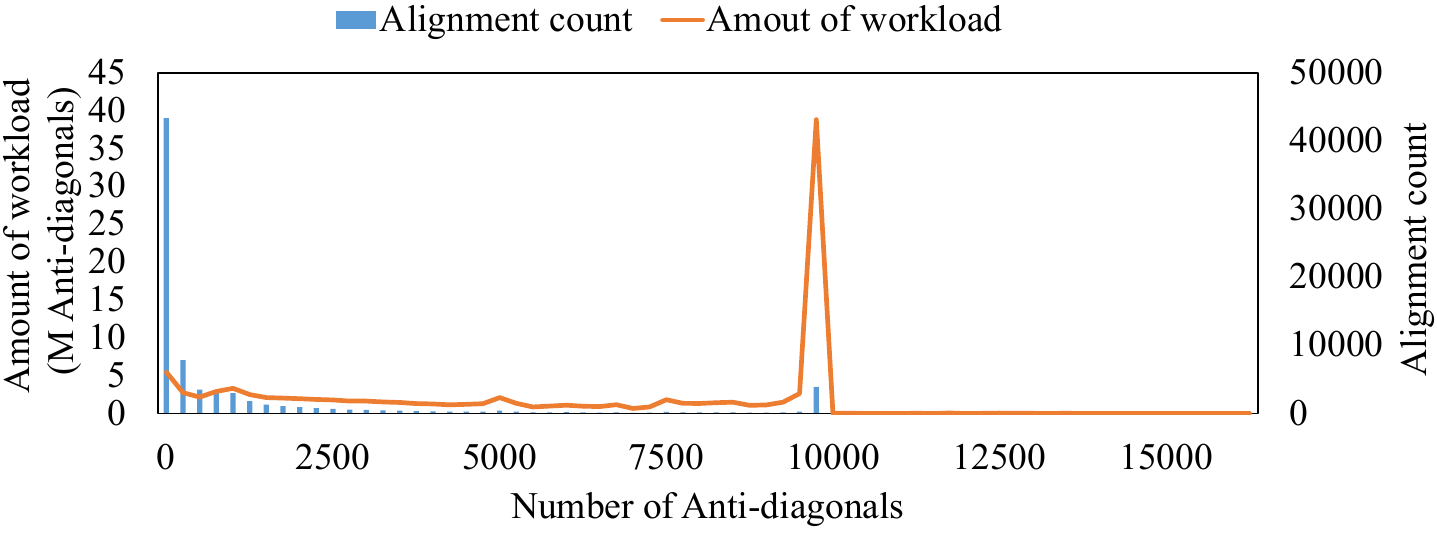}
    }
    \caption{A motivational study. \rev{(a) represents the execution times of the existing CPU-based algorithm and two naive GPU-based alignments}, and (b) represents the distribution of the accumulated workloads and alignment count for alignment tasks.}
    \label{fig:moti}
\end{figure*}

\subsection{Diagnosis of the Baseline Design}
\label{sec:moti:diagnosis}

In this section, 
we reveal four issues when orthogonally implementing the exact algorithm for guided alignment on the existing GPU-accelerated baseline design (\cref{sec:back:saloba}).

First, there is a direct problem of storing local maximum values (\cref{eq:localmax}) in the memory.
This causes pressure on the memory system and adds extra computation.

Second, 
there is a problem of \emph{run-ahead processing} as depicted in \cref{fig:saloba}(b). 
The termination condition requires computing the maximum value among the cells along each anti-diagonal.
However, during the horizontal progress of the threads, many anti-diagonals are not fully calculated, prohibiting the evaluation of termination conditions. 
For example, the termination condition at the red anti-diagonal line cannot be evaluated even after the entire chunk 1 is completed.
Therefore, when the termination is performed, a huge region is unnecessarily executed. 

Third, 
while using subwarps reduces external fragmentation, it is instead taxed with a heavy intra-warp workload imbalance.
This imbalance becomes larger and harder to predict, especially with the termination condition.

Last but not least, prior work does not consider inter-warp workload imbalances. 
Existing approaches assign tasks to warps in the order in which the input is given. 
This becomes a huge issue according to our study (\cref{sec:moti:exp}), which reveals several outliers that can cause a small portion of warps to handle all the heavy computation with long sequences.

\begin{figure*}
    \centering
    \includegraphics[width=\textwidth]{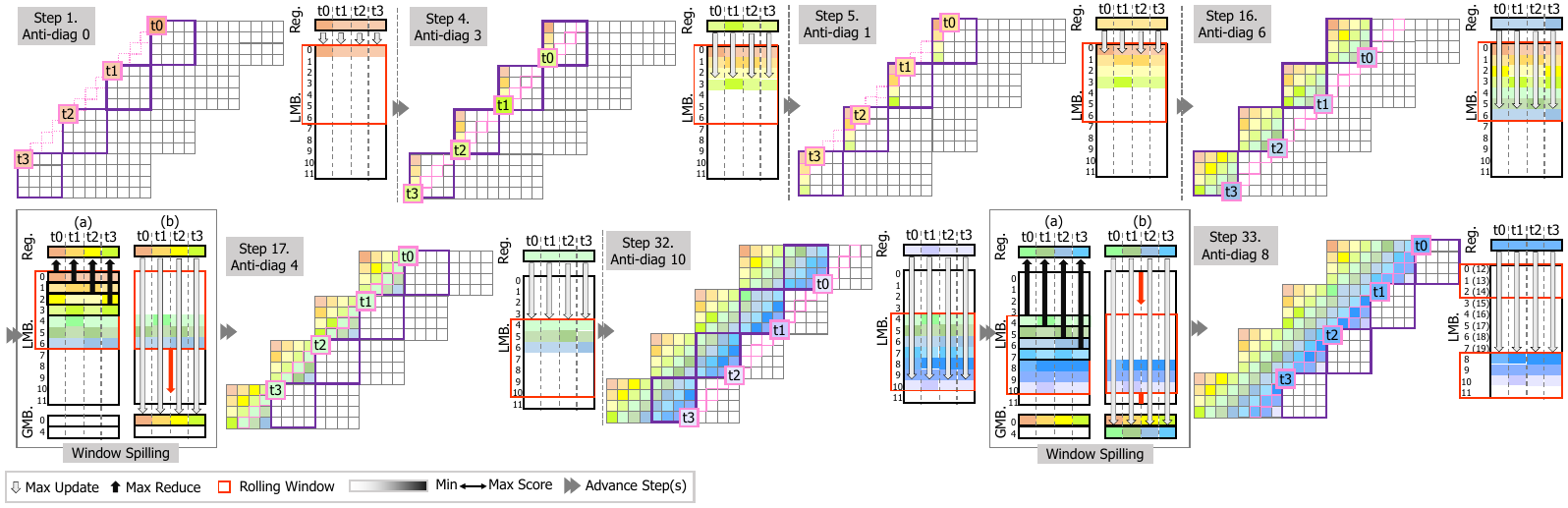}
    \caption{Tracking the anti-diagonal maximums with \mt. Cells on the same anti-diagonal are colored with the same hue.}
    \label{fig:max}
\end{figure*}

\subsection{Experimental Observation}
\label{sec:moti:exp}
In this section, we support our diagnosis with experimental observations.
\cref{fig:moti}(a) plots the execution time of a CPU-based alignment~\cite{minimap2} (`CPU'). 
\rev{We compared the performance with the baseline GPU acceleration described in \cref{sec:back:saloba} (\cite{saloba}) (`Baseline (Diff-Target)').  
Because this targets a different aligner from the baseline~\cite{minimap2}, we also extended it to support guiding techniques (`Baseline (MM2-Target)').}
\rev{As shown in \cref{fig:moti}(a),
the former version of baseline enjoys 5.3$\times$ geometric mean speedup from the CPU implementation.  
However, when extended to the latter, it becomes 2.0$\times$. }

One cause of the slowdown is the additional overhead from tracking the local maximum values, 
but another cause comes from the workload imbalance stemmed by the guiding techniques, as shown in \cref{fig:moti}(b).
The Y-axis represents the accumulated size of all workloads of tasks that fall into the range depicted in the X-axis.
Unlike most alignment tasks, certain tasks require significantly larger computations on the far right. 
This unique distribution works poorly with the baseline's design of assigning tasks to subwarps in the incoming order.  
As revealed in \cref{sec:moti:diagnosis}, 
if one subwarp is assigned a huge workload, this will cause both intra- and inter-warp imbalances that become the performance bottleneck.


\section{\thiswork Design}

\subsection{Tracking Local Maximums with \MT}
\label{sec:opt:max}

To test the termination condition, calculating each cell's score also updates the local maximum of the corresponding anti-diagonal (\cref{eq:localmax}).
The local maximum values are to be preserved until the entire anti-diagonal is processed and tested against the termination condition (\cref{eq:drop}).
As the threads progress horizontally, processing a chunk leaves many incomplete local maximums, requiring further processing. 
Thus, we must store the partial maximum values in the memory, which becomes a huge performance burden.

We devise a \mt approach to temporarily store partial maximums of calculated anti-diagonals in the shared memory and periodically spill them to the global memory, as in \cref{fig:max}.
To do this, we allocate a \emph{local maximum buffer (LMB)} in shared memory, organized in a 2-D table of $3\cdot block\_size \times num\_threads$.
In this example, we use a subwarp with 4 threads ($t0$ to $t3$) and blocks with 4x4 cells 
(requires $12\times 4$ table). 
The grid represents cells within the score table, and the purple boxes represent the anti-diagonal blocks being processed. 
Notice that each block spans over an equal set of seven anti-diagonals, and all threads are always on the same anti-diagonal depicted with pink.
The \mt comprises seven rows of the LMB (depicted red), representing the anti-diagonals in the current blocks' scope.
\Mt organizes each thread to keep its partial maximum values in its designated column of the window in LMB. 
When the window rolls down, the values for complete local max values are reduced and spilled to the \emph{global max buffer (GMB)} in the device memory as the following:  

\begin{enumerate}
    \item (Step 1) Each thread is processing the top-left cell of a block, being on the same anti-diagonal. They calculate the cell scores into registers.  
    The values are then written to the rolling window's first row in LMB, representing the first anti-diagonal's thread-local maximum values. 
    \item (Step 4) The threads proceed vertically inside a block. The first four processed cells belong to four different anti-diagonals, 
    storing the values on each corresponding row within the \mt. 
    \item (Step 5) Next, the threads are back to processing the second anti-diagonal.
    The threads compare and update the second thread-local max values in the rolling window. 
    \item (Step 16) All 4x4 cells of the block have been processed, which correspond to anti-diagonals $0\sim 6$.
    Note that anti-diagonals $0\sim 3$ are complete, while $4\sim 6$ are not.
    \item (\WS) After step 16, a spilling step follows.
    \begin{enumerate}
        \item Threads select one of the completed four rows in the LBM and read it to perform max-reduction using 
        $\mathit{\_\_reduce\_max\_sync}$ warp intrinsic. 
        Afterward, the four rows are cleared. 
        \item The reduced values are updated to the GMB. 
    \end{enumerate}
       \item (Step 17) Each thread progresses horizontally onto the next block. 
    The cells of this new block correspond to anti-diagonals $4\sim 10$.
    The rolling window also `rolls down' to target the next anti-diagonals. 
    \item (Step 32$\sim$33) The procedure repeats for the next blocks. 
\end{enumerate}
This consumes a reasonable amount of shared memory
while removing redundant memory accesses and utilizing coalescing. 
Note that the values from rows remaining in the window (e.g., ($4\sim 6$) in step 17) are re-accessed in the later steps. 
Additionally, in a special case where LMB is large enough to keep all the anti-diagonals, 
the spilling can be skipped (see \cref{sec:kernels:sd} for details).

\begin{figure}
    \centering
    \includegraphics[width=\columnwidth]{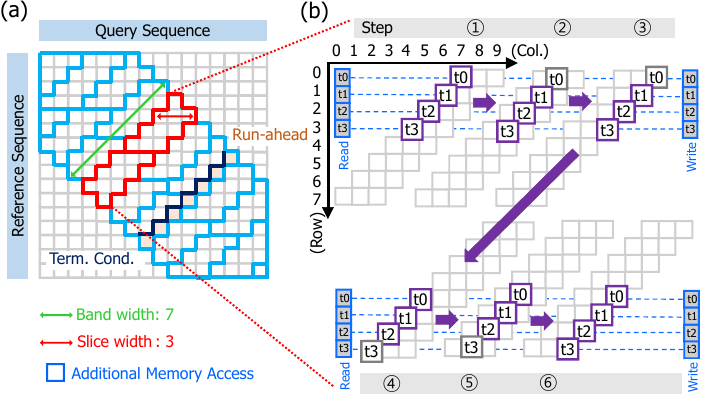}
    \caption{\Sd strategy. }
    \label{fig:kernels:slice}
\end{figure}

\begin{figure*}
    \centering
    \includegraphics[width=\textwidth]{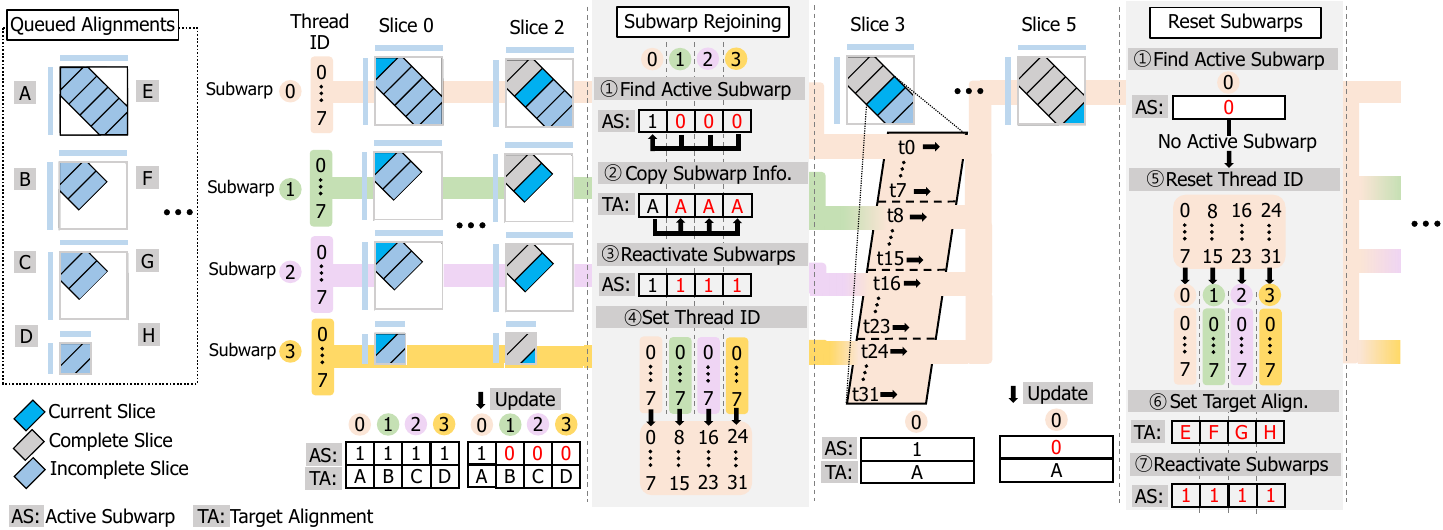}
    \caption{\Sr.}
    \label{fig:sr}
\end{figure*}

\subsection{\SD Strategy}
\label{sec:kernels:sd}
As exemplified in \cref{sec:moti:diagnosis} and \cref{sec:opt:max}, the horizontal-only progress strategy has several drawbacks on max tracking and run-ahead processing. 
To address these, we propose a novel tiling scheme named \sd, depicted in \cref{fig:kernels:slice}.
In this strategy, we partition the band into multiple \emph{slices} (red) along the anti-diagonal direction as shown in \cref{fig:kernels:slice}(a) with the width of $s$ (in the figure, 3 blocks).
Each slice is further horizontally partitioned into chunks whose number of rows is equal to the number of threads in a subwarp (in the figure, 4 blocks).
Within a slice (\cref{fig:kernels:slice}(b)), each thread within a subwarp computes blocks on the same row (\midcircled{1}$\sim$\midcircled{3}). 
Since the threads access the same row, they can keep the intermediate values using the registers. 
Then all threads in a subwarp move on to the next block to the next chunk to the bottom left (\midcircled{4}), and continue until the slice is completed (\midcircled{5}$\sim$\midcircled{6}).
When the entire slice is finished (\midcircled{6}), we check the termination condition for the completed anti-diagonals in the slice.

The proposed kernel enjoys the benefits of reduced run-ahead execution and the shared memory requirement of local max tracking. 
First, the amount of run-ahead execution size is greatly reduced. 
Unlike the baseline kernel, the run-ahead execution does not exceed $s \times Band\_width$ as again depicted with brown color in \cref{fig:kernels:slice}.
Second, it reduces the shared memory requirement for the \mt.
As mentioned in \cref{sec:opt:max}, if the number of entire anti-diagonals in a slice is small enough to fit in the LMB, this can eliminate the need for global memory accesses from \mt.

However, there is an interesting trade-off with these overhead reductions in memory access.
When a chunk in a slice starts/ends, it must read/write horizontal intermediate values (scores from $(i,j-1)$ in \cref{eq:sm-f}) for the following slices as in \cref{fig:kernels:slice}(b). 
Thus as $s$ decreases, the number of total slices within the score table increases, leading to more memory access. 
After some tuning effort (details in \cref{sec:eval:kernels}), we settled at $s=3$. 
Note that when $s$ is larger than the band width, the \sd kernel reduces to the baseline kernel, making \sd a generalization of the baseline.

\subsection{Reducing Warp Divergence with \SR}
\label{sec:kernels:sr}

\label{sec:opt:sr}

\begin{table*}[t]
    \centering
    \small
    \caption{Performance Models}
    \label{tab:Perf_model}
    \begin{tabular}{ll}
    \toprule
    Design & \qquad \qquad \qquad \qquad \qquad \qquad \qquad \quad Model \\ 
    \midrule
    Baseline & $\mathit{{M\widetilde{A}X}_{Warps}(MAX_{Subwarps}(Cells \hspace{3.8mm} \times({\frac{1}{Comp. TP}}+\frac{{AR_{Anti}\hspace{3.8mm}+AR_{Inter} \hspace{3.8mm}+AR_{Term}\hspace{3.8mm}}}{Mem. TP})))}$ \\ [8pt]
    $+$RW & $\mathit{M\widetilde{A}X_{Warps}(MAX_{Subwarps}(Cells \hspace{4mm} \times({\frac{1}{Comp. TP}}+{\frac{AR_{Anti}({\color{red}\downarrow}) + AR_{Inter} \hspace{3.8mm} + AR_{Term} \hspace{3.8mm}}{Mem. TP}})))}$ \\ [8pt]
    $+$RW$+$SD &$ \mathit{M\widetilde{A}X_{Warps}(MAX_{Subwarps}(Cells({\color{red}\downarrow})\times({\frac{1}{Comp. TP}}+{\frac{AR_{Anti}({\color{red}\downarrow}) + AR_{Inter}({\color{red}\uparrow}) + AR_{Term}({\color{red}\downarrow})}{Mem. TP}})))}$ \\ [8pt]
    $+$RW$+$SD$+$SR & $\mathit{{M\widetilde{A}X}_{Warps}({\color{red}{A\widetilde{V}G}}\hspace{0.5mm}_{Subwarps}(Cells \hspace{4mm} \times({\frac{1}{Comp. TP}}+{\frac{AR_{Anti}\hspace{3.8mm} + AR_{Inter} \hspace{3.8mm} + AR_{Term} \hspace{3.8mm}}{Mem. TP}})))}$ \\ [8pt]
    $+$RW$+$SD$+$SR$+$UB & $\mathit{{\color{red}{A\widetilde{V}G}}\hspace{0.5mm}_{Warps}({A\widetilde{V}G}\hspace{0.5mm}_{Subwarps}(Cells \hspace{4.2mm} \times({\frac{1}{Comp. TP}}+{\frac{AR_{Anti}\hspace{3.8mm} + AR_{Inter} \hspace{3.8mm} + AR_{Term} \hspace{3.8mm}}{Mem. TP}})))}$ \\ [8pt]
    \bottomrule
    \end{tabular}
\end{table*}
\Sr is a scheme designed to address the warp divergence from subwarps' workload imbalance. 
\cref{fig:sr} shows an example warp with four subwarps of eight threads each, where only the first subwarp is assigned a large alignment task. 
Because those subwarps belong to a single warp, this might cause significant underutilization.

To amend this issue, we design a novel scheme called \sr.
\Sr is a form of work stealing, but the novelty lies in that it is tightly coupled to the application such that it operates in a low overhead and fine-grained manner. 
In short, \sr waits until a working subwarp finishes a slice, and rejoins idling subwarps to form a larger subwarp. 
The detailed steps of \sr are as the following:

\begin{enumerate}
    \item (Slice 0) Each subwarp starts with the first slice of its alignment task. 
    Each subwarp maintains a flag in the shared memory for being active or not (Active Subwarp, AS) and the ID of said task (Target Alignment, TA).
    \item (Slice 2) After slice 2 is completed, tasks B, C, and D are finished. 
    The respective subwarps update AS to 0, which triggers \sr. 
    \item (\SR) works as the following:
    \begin{enumerate}
        \item The deactivated subwarps search AS to find an active subwarp (subwarp 0).
        \item The deactivated subwarps copy information from the active subwarp's task in TA.
        \item The subwarps set their AS flags back to active.
        \item Subwarps are merged by adjusting local thread IDs 
        using $\mathit{\_\_match\_any\_sync}$ warp intrinsic.
    \end{enumerate}
    \item (Slice 3) Subwarp 0 (now the entire warp) computes the remainder of slice 3 with all 32 threads. 
    \item After task A is completed, AS is updated.
    \item (Reset Subwarps) As no active subwarp remains according to AS, the subwarps are re-split to the original sizes and each fetches a new task. 
\end{enumerate}

The key of \sr is synchronizing the subwarps at slice boundaries. 
This indicates another trade-off with the slice width $s$, where a wide slice would cause longer subwarp idle time, and the contrary would cause too frequent overhead of checking for \sr feasibility.

\subsection{Workload Balancing with \UB}
\label{sec:kernels:ub}

\begin{figure}
    \centering
    \includegraphics[width=\columnwidth]{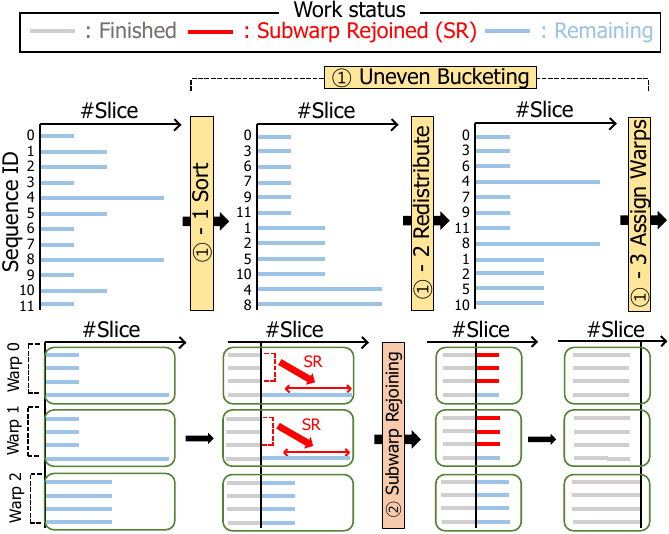}
    \caption{\Ub used with \sr.} 
    \label{fig:ub}
\end{figure}

\Ub is the last piece of \thiswork designed to reduce the inter-warp workload imbalance. 
In the baseline method, the alignment tasks are assigned to warps in an arbitrary order (i.e., sequential).  
However, in the existence of extra-long workloads, as depicted in \cref{fig:moti}(b), some warps will receive much longer workloads than others, which causes the performance to be dominated by the slowest warp.  

To prevent this from happening, we use \ub as illustrated in \cref{fig:ub} to distribute long sequences to different warps, flattening out the workload per warp. 
The method is straightforward to implement:
\begin{enumerate}
    \item Sort the given sequences to pick the longest $1/N$  sequences where $N$ is the number of subwarps per warp.
    \item Redistribute the sorted sequences such that one long sequence is assigned per warp. 
\end{enumerate}

\Ub owes to \sr on handling the uncertainty coming from the dynamic nature of the termination condition.
If the long task continues without termination, the large workload can be automatically redistributed with \sr.
If termination does occur, \sr can still reduce the overall execution time by rejoining the terminated subwarp to other active subwarps.


\subsection{Performance Modeling}

We present a simple performance model for the latency of each scheme of \thiswork in \cref{tab:Perf_model}. 
Within a subwarp, the latency is proportional to the total number of cells in the score table, simplified as below:
\begin{align}
    \mathit{\mathit{Cells} = \mathit{Antidiags} \times \mathit{Band\_width} + \mathit{Runahead}}.
\end{align}

\begin{sloppypar}
Then we model its corresponding cost for computation and memory access.
All cells in the scoreboard are processed at a certain  
computational throughput ($Comp. TP$), and require memory access at memory throughput ($Mem. TP$)).
There are three parameters that model the portion of cells that access memory: 
memory access ratio for storing the anti-diagonal max values ($\mathit{AR_{Anti}}$), managing intermediate values ($\mathit{AR_{Inter}}$), and checking the termination condition ($\mathit{AR_{Term}}$).
At the baseline, these ratios can be approximated as $1 \colon (1/8) \colon (1/\mathit{Band\_width})$, respectively.
Then, a combination of the subwarp latencies models the warp latency, and a combination of the warp latencies models the total latency.
\end{sloppypar}

In the baseline, the subwarps have no interaction, so  
the longest warp dominates the overall latency ($\mathit{MAX_{Subwarps}()}$).
Similarly, in the existence of extremely long queries, the longest warp will dominate the execution time ($\mathit{{M\widetilde{A}X}_{Warps}()}$) where $\mathit{M\widetilde{A}X}(\cdot)$ denotes a function dominated by maximum. 
 
\begin{sloppypar}
By applying \textbf{\mt (RW)}, the kernel can use shared memory to 
greatly reduce the number of memory accesses for anti-diagonal max tracking. 
This is depicted by reduced $AR_{Anti}$ in the model.
Next, by adding \textbf{\sd (SD)}, we can mainly reduce the amount of run-ahead processing, leading to an overall decrease in $Cells$. 
We can additionally decrease both $AR_{Anti}$ and $AR_{term}$ by using an optimal slice length.
As discussed in \cref{sec:kernels:sd}, it slightly increases $AR_{inter}$, but the benefit outweighs the penalty.
Applying \textbf{\sr (SR)} changes ${MAX}^{}_{Subwarps}()$ close to the average (${A\widetilde{V}G}_{Subwarps}()$) by letting subwarps help others within the warp.
Finally, \textbf{\ub (UB)} has the effect of reducing the straggler warps and changes the maximum dominated latency ${M\widetilde{A}X}_{Warps}()$ close to their average ${A\widetilde{V}G}_{Warps}()$.
\end{sloppypar}


\vspace{-2mm}

\section{Evaluation}
\label{sec:eval}

\subsection{Experimental Setup}
\label{sec:eval:setup}
We evaluated \thiswork on a server with NVIDIA RTX-A6000 GPU and AMD EPYC 7313P 16-Core Processor with 64 GB of RAM. It runs on Ubuntu 20.04 with CUDA version 11.7 and driver version 525.60.13.

For real-world datasets, one reference and nine query datasets were used. 
For reference, GRCh38~\cite{refsequence}, an up-to-date assembly of the human genome was used that has 3.1G base pairs (i.e., literals).
For the nine query datasets, we used data from the `Genome in a Bottle' project~\cite{genome-in-a-bottle}.
The datasets can be grouped into three categories by the sequencing techniques used to generate the dataset. 
The first category consists of HiFi HG (human genome) 005$\sim$007, which the PacBio HiFi~\cite{hifi} sequencing technique was used to obtain reads from the ChineseTrio.
The second category contains CLR 002$\sim$004 generated by the PacBio CLR~\cite{clr} sequencing technique from the AshkenazimTrio.
The last category holds reads from the same target, but the reads were extracted with the ONT~\cite{ont} sequencing technique.
We picked the first 50,000 reads from each dataset, and ran them through the pre-computing steps~\cite{minimap2} to obtain the final datasets for alignment. 
We mainly use Minimap2~\cite{minimap2} as the reference algorithm, 
but we also test the feasibility of \thiswork against BWA-MEM~\cite{bwamem}~(\cref{sec:eval:bwamem}).
We used Minimap2's preset parameters for each dataset category.  

\subsection{Baselines}

The GPU-based baselines chosen for the evaluation are as follows:

\begin{itemize}
\item 
\textbf{Manymap~\cite{manymap}} is a GPU alignment based on Minimap2. 
However, Manymap applies an inexact interpretation of the termination condition.
In addition, it allows aligning only one sequence at a time, so we fixed it to accept multiple different reads in parallel using CUDA streams.

 \item 
 \textbf{GASAL2~\cite{gasal2}} is a GPU alignment with input packing and inter-query parallelism. 
 Out of the multiple kernels implemented, we use the banding kernel in GASAL2.

 \item 
 \textbf{SALoBa~\cite{saloba}} utilizes intra-query parallelism. 
 We applied banding heuristic that gives further speedup, similar to GASAL2.

 \item 
 \textbf{LOGAN~\cite{zeni2020logan}} is an algorithm that implements its own guiding algorithm. 
 \rev{It adjusts the band width during score table filling 
 after calculating each anti-diagonal as a form of guided alignment.}

\end{itemize}
For fair comparison, we measure the performance of the baselines in two versions. 
\rev{One is measured in the version that they originally target (Diff-Target), and another is extended with faithful optimization efforts to provide output equal to the reference algorithm (MM2-Target).}
This represents a scenario where those libraries are used to accelerate Minimap2 in the field. 
We fixed the provided termination condition for Manymap and newly implemented it in GASAL2/SALoBa. 
\rev{Since LOGAN targets a different guiding algorithm, 
we only report the unmodified performance for LOGAN.}
\rev{Note that Manymap is the only baseline intended as a direct replacement of Minimap2's algorithm. 
Others target different algorithms, thus not implementing all of the heuristics we focus on.}

\subsection{Performance Comparison}
\label{sec:eval:perf}
\begin{figure}
\includegraphics[width=\columnwidth]{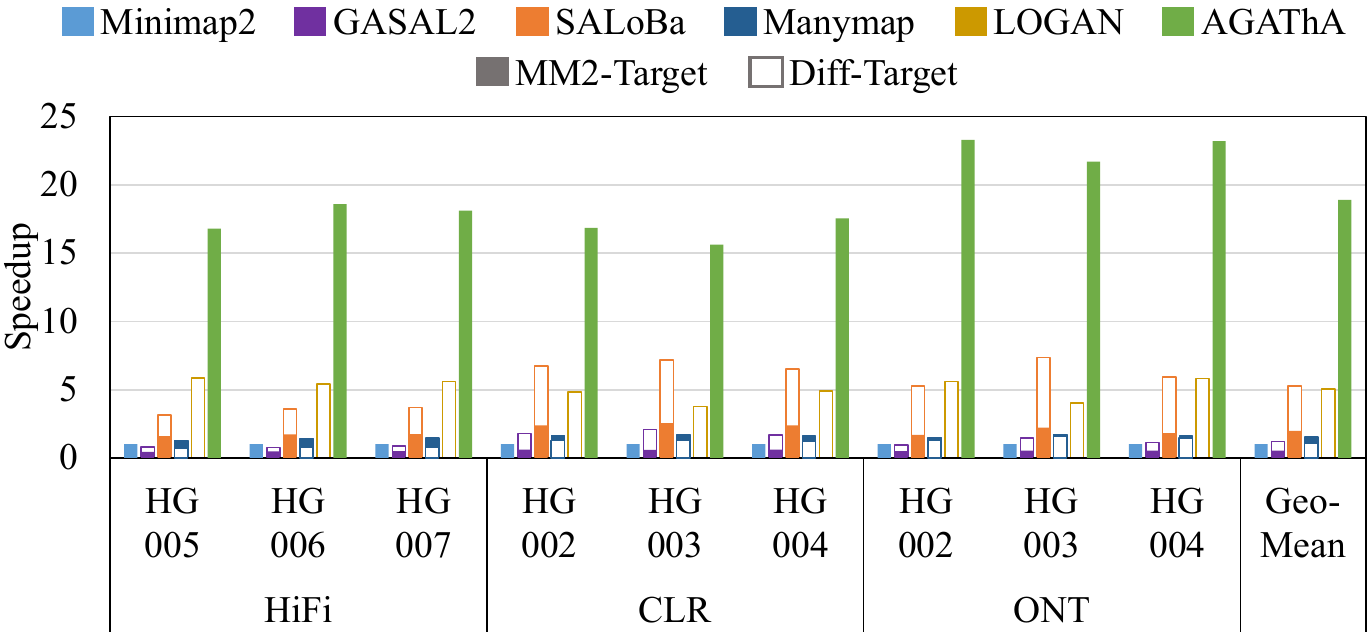}
\caption{\rev{Performance comparison.}}
\label{fig:eval:perf}
\end{figure}

In this section, we evaluate the performance of \thiswork using real-world sequencing datasets.
\cref{fig:eval:perf} reveals the performance comparisons of \thiswork against the baselines, normalized to the performance of Minimap2 on a CPU.
\rev{For the baselines, the blank bars represent the speed of the originally targeted algorithms (Diff-Target), while the solid bars represent the speed of implementations with the reference guiding algorithm as their target (MM2-Target).}

\rev{Against all baselines, \thiswork clearly outperforms them significantly.}
Over Minimap2, \thiswork showed 18.8$\times$ geometric mean speedup. 
While \thiswork shows more than 16$\times$ speedup for all datasets, it was the largest on the HiFi dataset (22.6$\times$). 

\rev{Between the Minimap2-targeted baselines, SALoBa was the fastest but was 9.6\x\ slower than \thiswork.}
Next, Manymap and GASAL2 were 12.1$\times$ and 36.6$\times$ slower than \thiswork, respectively.
\rev{This version of GASAL2 was even slower than Minimap2.
Also, Manymap is the only version that benefits (albeit slightly) from implementing the guided alignment algorithm.}
We suspect that this is from the fact that Manymap is already filling the score table by computing each anti-diagonal, removing the run-ahead processing entirely. 

\rev{Among baselines that do not target Minimap2, SALoBa was also the fastest but was still 3.6$\times$ slower than \thiswork.}
SALoBa's speedup comes from albeit naive, banding reducing the computational workload, and other GPU acceleration techniques such as subwarps were used. 
\rev{Additionally, LOGAN's performance closely follows SALoBa. 
One cause could be that LOGAN maintains a gap score that is less expensive in both computation and memory.} 
Manymap was the slowest with 1.1$\times$ geometric mean speedup.

\begin{figure}
\includegraphics[width=\columnwidth]{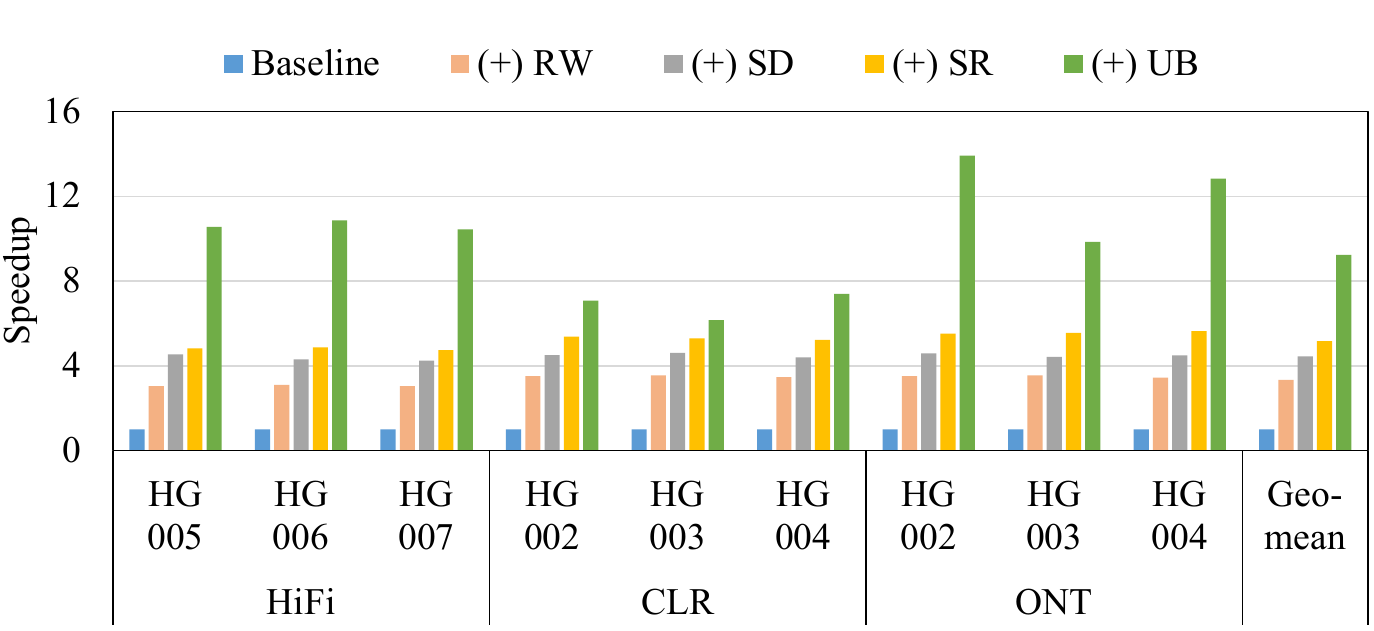}
\caption{Ablation study.}
\label{fig:eval:ab}
\end{figure}

\subsection{Ablation Study}
\label{sec:eval:ab}
To demonstrate the advantage of how \thiswork achieves its speedup, we conduct an ablation study 
as in \cref{fig:eval:ab}.
\rev{For the baseline, we use the naive exact implementation of the guiding algorithm, without the proposed techniques.}
By using \mt for max tracking (RW), \thiswork already achieves an average of 3.1$\times$ on the HiFi datasets and 3.5$\times$ on the rest. 
This speedup comes from decreasing the number of global memory accesses and using max-reduce to parallelize calculating the anti-diagonal maximum.
\Sd (SD) further optimizes \mt by additional speedup of 1.4$\times$ for the HiFi datasets, and 1.3$\times$ for the other two datasets, respectively. 
It optimizes \mt by reducing the run-ahead execution and only using an essential amount of shared memory to reduce global memory access.  
In addition, \sr (SR) increases speedup by an additional 1.1$\times$ for the HiFi datasets, 1.2$\times$ for the rest.
It removes the warp divergence coming from subwarps receiving different-sized workloads. 
Finally, \ub (UB) boosts the alignment performance by approximately 2.2$\times$ more on the HiFi and ONT datasets, and 1.3$\times$ more on CLR datasets. 
\Ub spreads the few extremely large workloads to multiple warps to solve the inter-warp imbalance problem with the aid of \sr.

\begin{figure}
\includegraphics[width=\columnwidth]{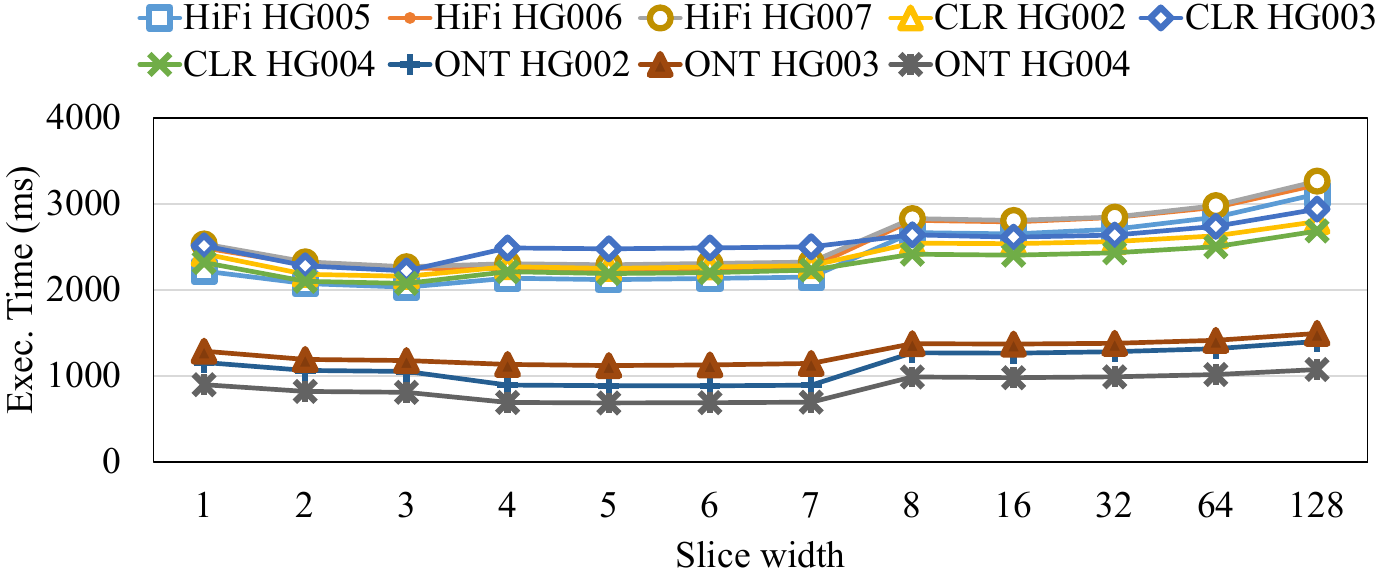}
\caption{Slice width sensitivity study.}
\label{fig:eval:slicewidth}
\end{figure}

\subsection{Sensitivity Study on Slice Width} 
\label{sec:eval:kernels}

The slice width is an important parameter that can highly affect the performance of the \sd kernel. 
In \cref{fig:eval:slicewidth}, we change the slice width from 1 up to 128 blocks. 
There is an overall decreasing trend from 1 to 4, flattening around 5 to 16, and an increasing trend as the slice width gets larger. 
The first decrease comes from reducing the number of memory accesses for intermediate values, and the increasing trend at the end comes from the growing amount of run-ahead processing. 
Among this overall trend, we can see small jumps after slice widths 3 and 7. 
This is because it is possible to use bitwise \& operation with these widths instead of modulo operation which is known to be slow on GPUs. 
While there were both similar speedups at widths 3 and 7, we chose 3 as our main target. 
This is because it uses less shared memory (and saves space for implementing \sr), and the speedup for the longer datasets HiFi and CLR were higher for width 3. 
The upper hand of width 3 on long datasets 
stems from the fact that the effect of memory access for anti-diagonal max scores and run-ahead execution is more important than short datasets.

\subsection{ \SR and \UB}
\label{sec:eval:sr}

\begin{figure}
\includegraphics[width=\columnwidth]
{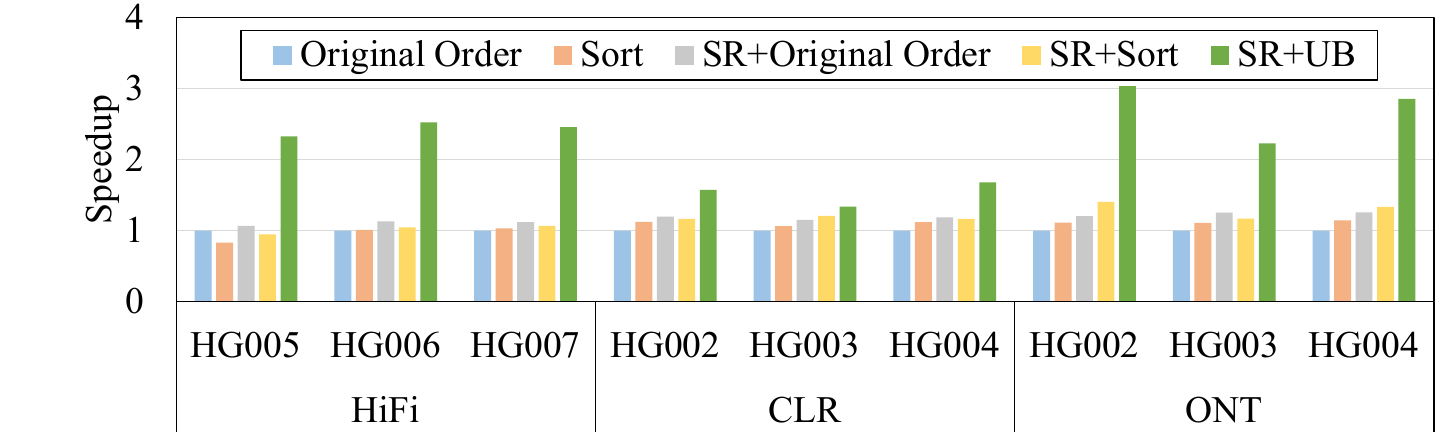}
\caption{Effect of the workload balancing techniques.}
\label{fig:eval:sr_speed}
\end{figure}

In this section, we study the impact of \sr and \ub by observing the speedup and distribution of allocated workload per thread. 
We compare them with sorting, which we believe is simple and intuitive for reducing workload imbalances.
\cref{fig:eval:sr_speed} shows the performance comparisons, where the `Original order' represents \thiswork only with \mt and \sd applied.
If we apply sorting to the workloads by the number of anti-diagonals, there was approximately 1.06$\times$ speedup in geometric mean for the entire dataset (`Sort'). 
This is from the reduced warp divergence by allocating similar-sized score tables to the subwarps within the same warp. 
However, applying \sr to the baseline (`SR + Original Order') shows a speedup of 1.17$\times$.
The result shows that \sr is more effective in reducing warp divergence than sorting.
This is because sorting cannot adapt to the dynamic behavior of termination, while \sr can.

If we apply \sr to the sorted sequences (`SR + Sort'), the speedup is still 1.17$\times$ in geometric mean, similar to only applying \sr. 
Finally, with \sr and \ub, the speedup is 2.22$\times$ on average for all datasets. 
This shows how \sr and \ub working in unison is the best approach in reducing workload divergence both within and between warps.

\begin{figure}
\includegraphics[width=\columnwidth]
{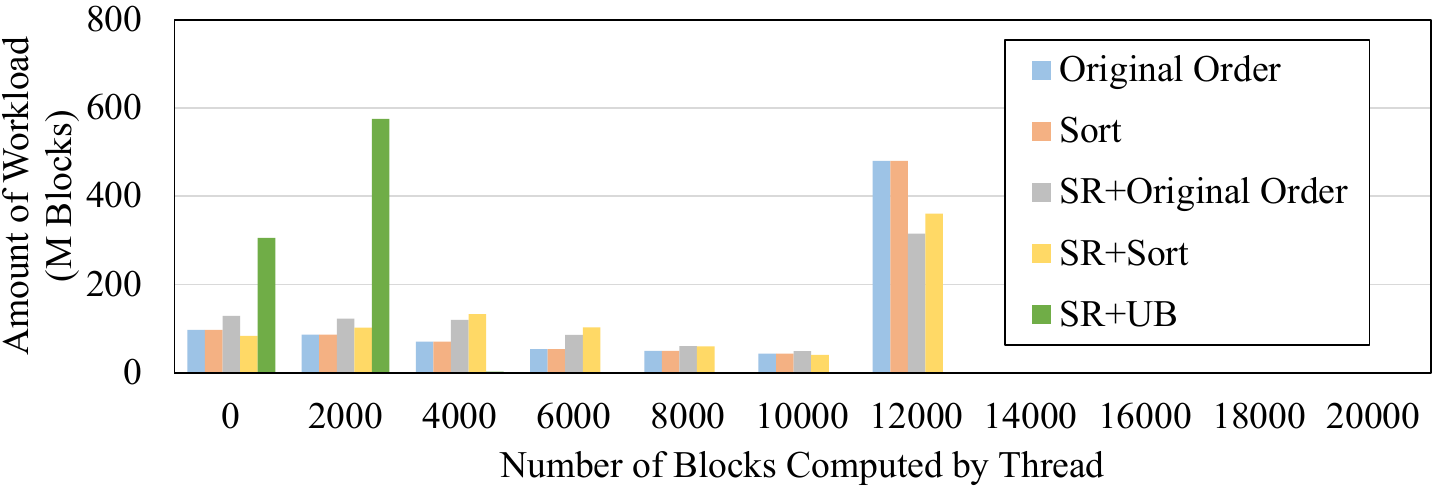}
\caption{Workload distribution from workload balancing.} 
\label{fig:eval:sr_work}
\end{figure}

\cref{fig:eval:sr_work} shows how the combination of \sr and \ub redistributes the workload evenly to the subwarps. 
The Y axis shows the accumulated number of blocks, representing the amount of workload.
On the `Original Order', much of the work is performed on subwarps whose number of initially assigned blocks per thread is around 12,000.
While `SR+Original Order' and `SR+Sort' do reduce the work from the heavily loaded subwarps, their impact is limited. 
Contrarily, we can see that \sr and \ub together shift the entire graph to the left, which represents how the large amounts of workloads that were allocated to a few subwarps are spread out to many other subwarps. 
This shows the underlying mechanism of how \ub is such a powerful acceleration tool with our data distribution. 

\begin{figure}
    \centering
    \includegraphics[width=\columnwidth]{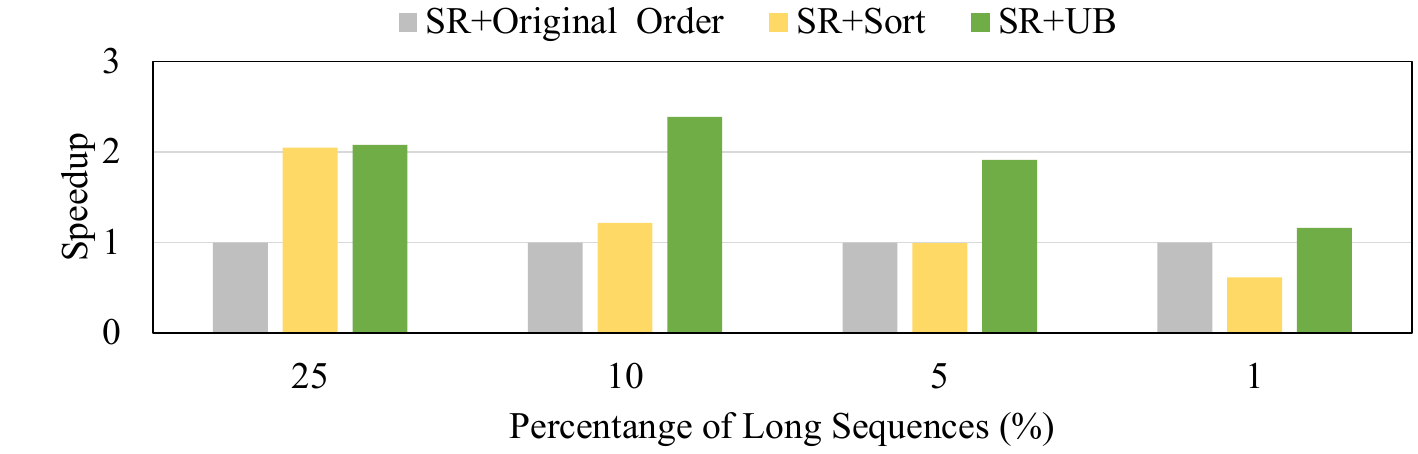}
    \caption{Performance on datasets with different long sequence percentages.}
    \label{fig:eval:long_ratio}
\end{figure}

To better understand why \ub shows significant speedup, we compare the performance of \ub on generated datasets in \cref{fig:eval:long_ratio}.
We generated datasets by varying the percentage of long sequences (4096 bp) against short sequences (128 bp).
To isolate the impact of \ub, we used `SR+Original Order' as the baseline 
and compared `SR+Sort' and 'SR+UB'. 
Even with a similar speedup at 25\%, \ub always outperformed sorting for every 
percentage, showing the highest speedup of 2.39$\times$ at 10\%. 
Notably, the speedup of sorting peaks at 25\% and continues to drop as the percentage decreases, even becoming slower than the original ordering by 0.61$\times$. 
The main reason for this drastic slowdown is that a few warps with long sequences become the main bottleneck.
Contrarily, \ub is still faster than the original ordering because it distributes the long sequences to different warps and can fully utilize \sr.
Note that the percentage of alignments on the far right peak in \cref{fig:moti}(b) ranged between 5$\sim$20\% for all datasets.

\subsection{Sensitivity Study on Subwarp Size}
\label{sec:eval:subwarp}

\begin{figure}
\includegraphics[width=\columnwidth]{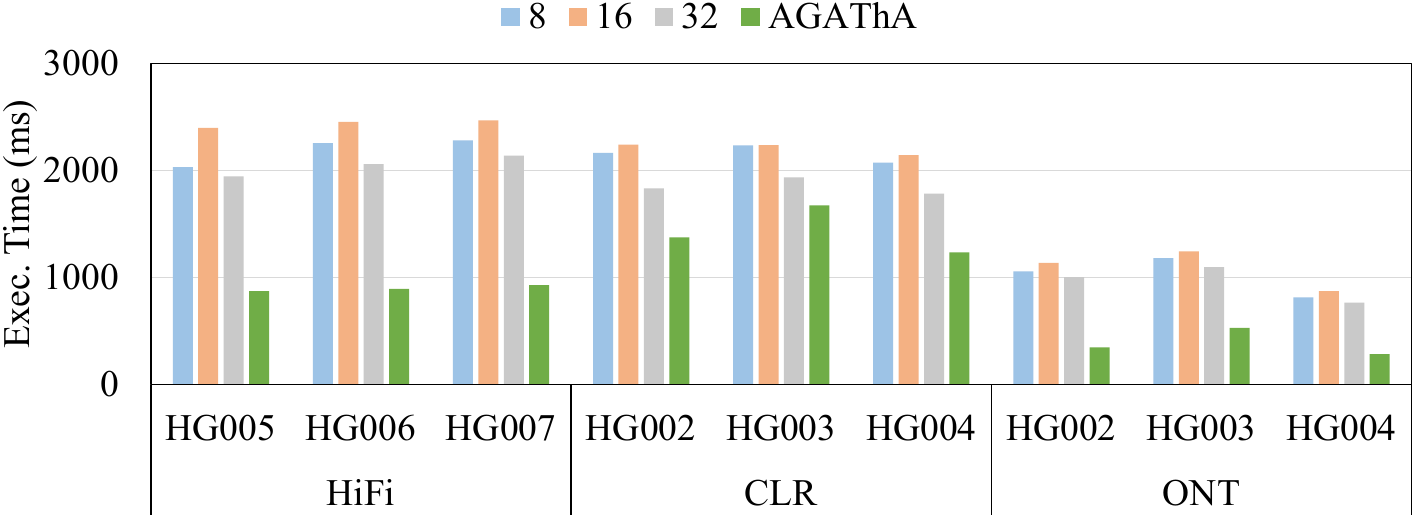}
\caption{Sensitivity study on subwarp size.}
\label{fig:eval:subwarp}
\end{figure}

We conducted a sensitivity study on the effect of different subwarp sizes on execution time.
\cref{fig:eval:subwarp} shows the execution time for different subwarp sizes 8, 16, and 32 (full warp) compared to the final \thiswork.
Due to the nature of subwarps, 
warp divergence increases as the subwarp size decreases. 
One could think it would be faster to use the full warp per alignment instead of subwarps. 
For the kernel with \mt and \sd only, it is true that using the full warp is faster than using subwarps, on average 10\%.
However, this is easily outpaced by the final version of \thiswork, which uses designs such as \sr and \ub that must have a subwarp to be implemented. 
Additionally, there are slowdowns at 16 threads per subwarp. 
This can be explained by the increased amount of idling threads at the start and end of a chunk with the remaining warp divergence from the usage of subwarps.

\subsection{Hardware Flexibility of \thiswork}

\begin{figure}
\includegraphics[width=\columnwidth]{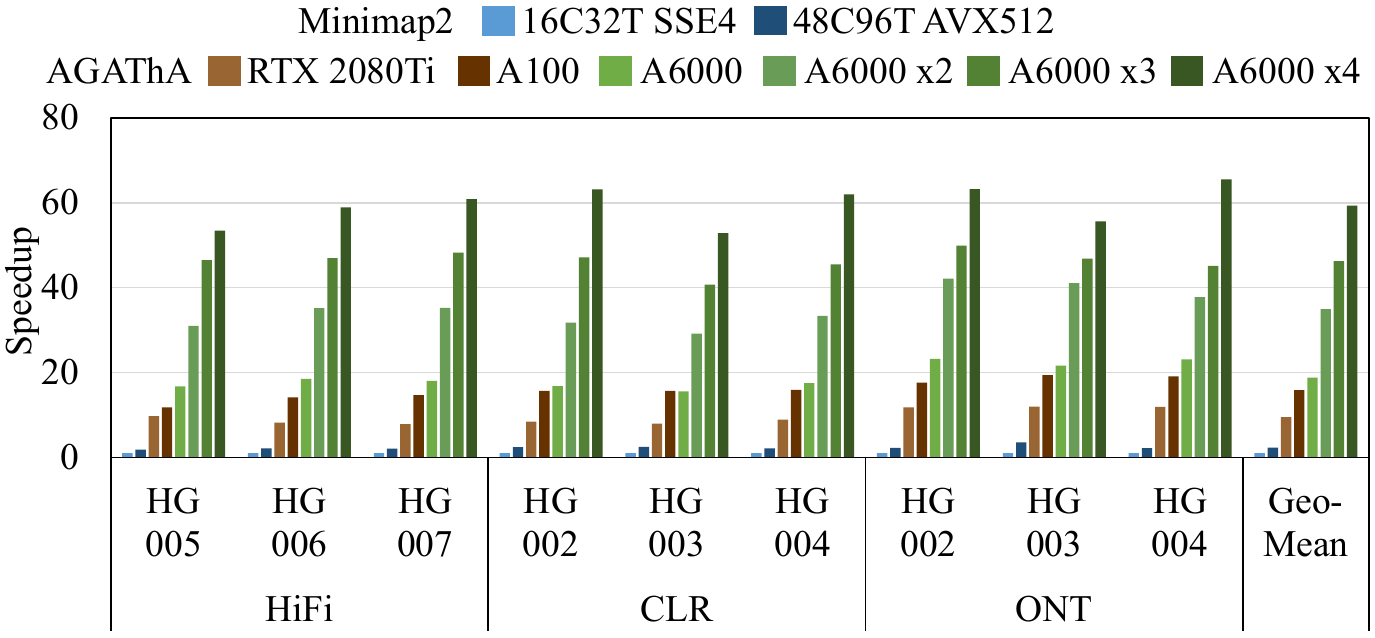}
\caption{Hardware flexibility of \thiswork.}
\label{fig:eval:diffhw}
\end{figure}

In the present section, we demonstrate the performance of \thiswork on various hardware environments in \cref{fig:eval:diffhw}. 

\textbf{Comparison with a Stronger CPU Baseline.}
In addition to the default CPU baseline implemented with SSE4.1 support that runs on a 16-core 32-thread processor (`16C32T SSE4'), we also experimented with a stronger CPU baseline with AVX512 support using optimized implementation from~\cite{minimap2avx} on a 48-core 96-thread environment (2 $\times$ Xeon Gold 6442Y Processor, `48C96T AVX512').
Overall, the stronger baseline was 2.30$\times$ faster in geometric mean compared to the default one.
Even with the stronger CPU baseline, \thiswork still showed a significant 8.19$\times$ geometric mean speedup with a single GPU.

\textbf{Sensitivity on GPU Types.}
We also tested \thiswork with RTX 2080Ti and A100 to verify the applicability on other GPUs. 
As RTX 2080Ti does not support 
warp-reduce functions from recent GPUs, we replaced them with shared memory access.
\thiswork provided stable 9.49$\times$ and 15.84$\times$ speedup over the CPU baseline in geometric mean, respectively.
Although A100 is perceived as a higher grade than A6000, A6000 performs better due to having a larger cuda core count.

\textbf{Scalability on Number of GPUs.}
Additionally, we extend \thiswork to multiple GPUs by distributing equal numbers of alignment tasks to each GPU and making each GPU process them.
\thiswork showed almost linear scalability and achieved 59.38$\times$ geometric mean speedup with four GPUs over the CPU baseline, which is close to linear compared to 18.83$\times$ speedup from a single GPU.
We expect \thiswork to perform even better by supporting workload balancing among different GPUs similar to \ub.

\subsection{Applying \thiswork to BWA-MEM}
\label{sec:eval:bwamem}

\begin{figure}
\includegraphics[width=\columnwidth]{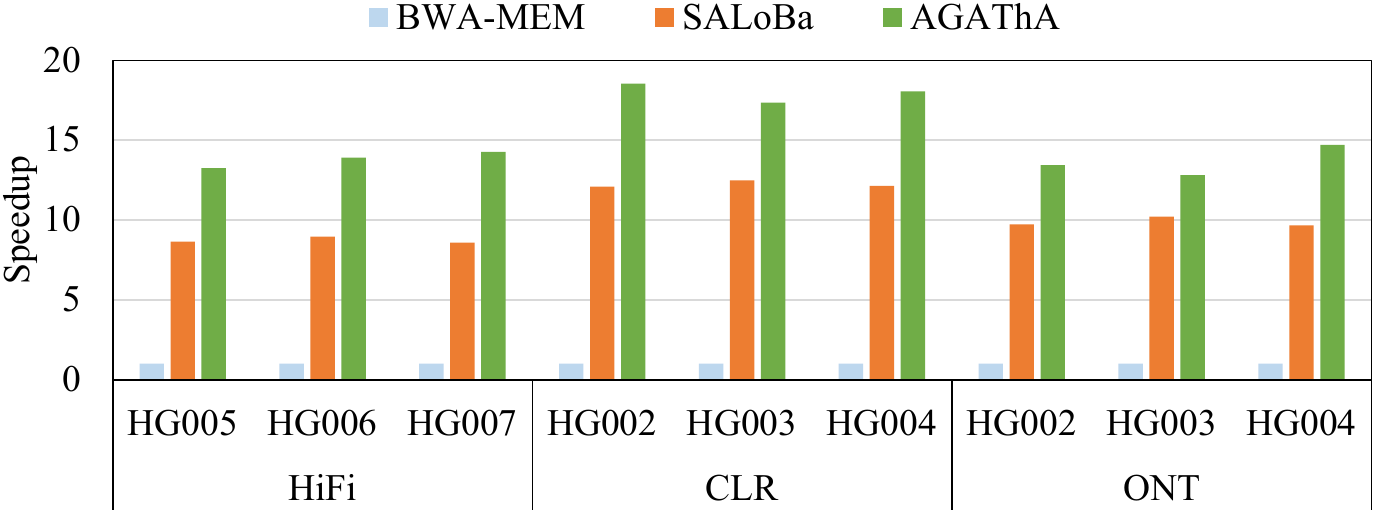}
\caption{Result of applying \thiswork to BWA-MEM.}
\label{fig:eval:bwa}
\end{figure}

Finally, we evaluate \thiswork on BWA-MEM~\cite{bwamem}, an algorithm for older generations that is still widely used. 
We demonstrate that our schemes can be applied to guided alignment algorithms other than Minimap2.
We compare the runtime of BWA-MEM on CPU with SALoBa and \thiswork by applying BWA-MEM's guided alignment.
We chose SALoBa, the fastest baseline from \cref{sec:eval:perf}.
\cref{fig:eval:bwa} shows that
\thiswork also has the speedup gap above SALoBa. 
The speed gap is smaller than on Minimap2 mainly due to the default band width and termination threshold being significantly smaller, effectively reducing the amount and imbalance of task workloads. 
However, \thiswork still achieves a significant speedup of 15$\times$ compared to BWA-MEM on a CPU.

    
\section{Discussion}
\textbf{Applying DPX to \thiswork.}
The new NVIDIA Hopper architecture~\cite{hopper} introduces a new instruction called Dynamic Programming extension (DPX)~\cite{dpx}.
It can accept two or three integers and compute ReLU or min/max comparison and is said to accelerate sequence alignments up to 4.4$\times$~\cite{dpxsw}.
We expect DPX to be seamlessly integrated into our kernel and help \thiswork thrive even more because DPX enhances the computational speed, 
and \thiswork mainly addresses the issues from memory bandwidth bottleneck.

\textbf{Different Bucketing Parameters.}
While \ub largely contributes to our speedup, we see more potential in this design.
For example, if we predict exactly when the termination condition is met before execution, then the kernel could remove most of the remaining workload imbalance.
We would like to explore this possibility in future work.


\section{Related Work}

\textbf{GPU Accelerations of Sequence Alignments.}
There is a line of work on accelerating sequence alignment with GPUs, using OpenGL library~\cite{liu2006gpu, liu2006bio}, or earlier version of CUDA~\cite{cudasw1, cudasw2, cudasw3, gswabe, gpu-pairalign}.
In addition, some target single-pair alignment and utilize intra-query parallelism~\cite{cudalign1, cudalign2, cudalign3, cudalign4}.
To be integrated into the read mapping algorithm, multiple alignments have to be performed on relatively shorter partial pieces of sequence. 
SOAP~\cite{soap3} and CUSHAW family~\cite{cushaw1,cushaw2,cushaw3} are early ones on such approach, and GASAL2~\cite{gasal2} introduces on-GPU input packing.
SALoBa~\cite{saloba} further optimizes on GPU memory hierarchy.
\rev{However, while they are compared with existing reference algorithms~\cite{minimap2, bwamem}, none have implemented the exact guiding algorithms.}

\textbf{Guided Alignment.}
Minimap2 and BWA-MEM are regarded as golden standards, but they are not the only algorithms with a guided dynamic programming approach.
For example, banding is implemented on other variant algorithms~\cite{minimap2avx, sadasivan2023accelerating}.
On GPUs, \cite{korpar2013sw} is a GPU-based alignment library with banding, 
and F5C~\cite{f5c} adaptively decides the direction of the band for every iteration. 
The termination condition has been first introduced in Blast~\cite{blast-xdrop}, called \textit{X-Drop}. 
However, it was found to penalize single long gaps too much. Because of this, it has been amended to be \emph{Z-drop} on later work as in \cref{eq:drop} by Minimap2~\cite{minimap2}.
It has been accelerated using SIMD instructions~\cite{minimap2avx}, but not on GPUs.

\textbf{Balancing Workload on GPUs.}
Workload balancing for GPUs is a classic problem studied for decades \cite{tzeng2010task, zhang2010streamlining, steinberger2012softshell, gupta2012study, chatterjee2013dynamic, steinberger2014whippletree, khorasani2015efficient}.
\cite{wu2015enabling} exploits software-level control of scheduling on streaming multiprocessors.
\cite{khorasani2016eliminating} suggests 
dynamic task-to-thread assignment,
and \cite{zheng2017versapipe} provides a pipelined parallel programming framework on GPU 
for job scheduling.
\thiswork also applies a type of workload balancing, but does so in deep relation with the algorithm and the observed real-world data distribution.

\section{Conclusion}
We propose \thiswork, a GPU-accelerated sequence alignment software. 
\rev{It is the first exact GPU implementation of the guided sequence alignment algorithm.}
To address the challenges the guiding step imposes, 
we propose a new method to track anti-diagonal maximums, a tiling strategy, a dynamic workload balancing, and a workload distribution strategy. 
The evaluation shows that \thiswork significantly outperforms the baselines. 

\begin{acks}

This work was supported by Samsung Advanced Institute of Technology, Samsung Electronics Co., Ltd (IO230223-05124-01),
the National Research Foundation of Korea (NRF) grant funded by the Korean government (MSIT) (2022R1C1C1011307, 2022R1C1C1008131),
and Institute of Information \& communications Technology Planning \& Evaluation (IITP) under the artificial intelligence semiconductor support program to nurture the best talents (IITP-2023-RS-2023-00256081) grant funded by the Korean government (MSIT). 
\end{acks}

\section*{Data Availability Statement}
The artifact of \thiswork is available at~\cite{agatha}.
The most recent version is kept updated at \url{https://github.com/readwrite112/AGAThA}.
The artifact contains a sample dataset, the whole code of \thiswork, and some evaluation-related code.
For the details of the artifact, please refer to \cref{sec:appendix}.

\newpage
\balance 

\bibliography{references}


\begin{thebibliography}{59}


\ifx \showCODEN    \undefined \def \showCODEN     #1{\unskip}     \fi
\ifx \showDOI      \undefined \def \showDOI       #1{#1}\fi
\ifx \showISBNx    \undefined \def \showISBNx     #1{\unskip}     \fi
\ifx \showISBNxiii \undefined \def \showISBNxiii  #1{\unskip}     \fi
\ifx \showISSN     \undefined \def \showISSN      #1{\unskip}     \fi
\ifx \showLCCN     \undefined \def \showLCCN      #1{\unskip}     \fi
\ifx \shownote     \undefined \def \shownote      #1{#1}          \fi
\ifx \showarticletitle \undefined \def \showarticletitle #1{#1}   \fi
\ifx \showURL      \undefined \def \showURL       {\relax}        \fi
\providecommand\bibfield[2]{#2}
\providecommand\bibinfo[2]{#2}
\providecommand\natexlab[1]{#1}
\providecommand\showeprint[2][]{arXiv:#2}

\bibitem[Ahmed et~al\mbox{.}(2019)]%
        {gasal2}
\bibfield{author}{\bibinfo{person}{Nauman Ahmed}, \bibinfo{person}{Jonathan L{\'e}vy}, \bibinfo{person}{Shanshan Ren}, \bibinfo{person}{Hamid Mushtaq}, \bibinfo{person}{Koen Bertels}, {and} \bibinfo{person}{Zaid Al-Ars}.} \bibinfo{year}{2019}\natexlab{}.
\newblock \showarticletitle{{GASAL2}: A {GPU} accelerated sequence alignment library for high-throughput {NGS} data}.
\newblock \bibinfo{journal}{\emph{BMC bioinformatics}}  \bibinfo{volume}{20} (\bibinfo{year}{2019}), \bibinfo{pages}{1--20}.
\newblock


\bibitem[Alser et~al\mbox{.}(2022)]%
        {alser2022molecules}
\bibfield{author}{\bibinfo{person}{Mohammed Alser}, \bibinfo{person}{Joel Lindegger}, \bibinfo{person}{Can Firtina}, \bibinfo{person}{Nour Almadhoun}, \bibinfo{person}{Haiyu Mao}, \bibinfo{person}{Gagandeep Singh}, \bibinfo{person}{Juan Gomez-Luna}, {and} \bibinfo{person}{Onur Mutlu}.} \bibinfo{year}{2022}\natexlab{}.
\newblock \showarticletitle{From molecules to genomic variations: Accelerating genome analysis via intelligent algorithms and architectures}.
\newblock \bibinfo{journal}{\emph{Computational and Structural Biotechnology Journal}}  \bibinfo{volume}{20} (\bibinfo{year}{2022}), \bibinfo{pages}{4579--4599}.
\newblock


\bibitem[Alser et~al\mbox{.}(2021)]%
        {alser2021technology}
\bibfield{author}{\bibinfo{person}{Mohammed Alser}, \bibinfo{person}{Jeremy Rotman}, \bibinfo{person}{Dhrithi Deshpande}, \bibinfo{person}{Kodi Taraszka}, \bibinfo{person}{Huwenbo Shi}, \bibinfo{person}{Pelin~Icer Baykal}, \bibinfo{person}{Harry~Taegyun Yang}, \bibinfo{person}{Victor Xue}, \bibinfo{person}{Sergey Knyazev}, \bibinfo{person}{Benjamin~D Singer}, {et~al\mbox{.}}} \bibinfo{year}{2021}\natexlab{}.
\newblock \showarticletitle{Technology dictates algorithms: recent developments in read alignment}.
\newblock \bibinfo{journal}{\emph{Genome biology}} \bibinfo{volume}{22}, \bibinfo{number}{1} (\bibinfo{year}{2021}), \bibinfo{pages}{249}.
\newblock


\bibitem[Altschul et~al\mbox{.}(1997)]%
        {blast-xdrop}
\bibfield{author}{\bibinfo{person}{Stephen~F Altschul}, \bibinfo{person}{Thomas~L Madden}, \bibinfo{person}{Alejandro~A Sch{\"a}ffer}, \bibinfo{person}{Jinghui Zhang}, \bibinfo{person}{Zheng Zhang}, \bibinfo{person}{Webb Miller}, {and} \bibinfo{person}{David~J Lipman}.} \bibinfo{year}{1997}\natexlab{}.
\newblock \showarticletitle{Gapped BLAST and PSI-BLAST: a new generation of protein database search programs}.
\newblock \bibinfo{journal}{\emph{Nucleic acids research}} \bibinfo{volume}{25}, \bibinfo{number}{17} (\bibinfo{year}{1997}), \bibinfo{pages}{3389--3402}.
\newblock


\bibitem[Awan et~al\mbox{.}(2020)]%
        {adept}
\bibfield{author}{\bibinfo{person}{Muaaz~G Awan}, \bibinfo{person}{Jack Deslippe}, \bibinfo{person}{Aydin Buluc}, \bibinfo{person}{Oguz Selvitopi}, \bibinfo{person}{Steven Hofmeyr}, \bibinfo{person}{Leonid Oliker}, {and} \bibinfo{person}{Katherine Yelick}.} \bibinfo{year}{2020}\natexlab{}.
\newblock \showarticletitle{{ADEPT}: A domain independent sequence alignment strategy for {GPU} architectures}.
\newblock \bibinfo{journal}{\emph{BMC bioinformatics}} \bibinfo{volume}{21}, \bibinfo{number}{1} (\bibinfo{year}{2020}), \bibinfo{pages}{1--29}.
\newblock


\bibitem[Blazewicz et~al\mbox{.}(2011)]%
        {gpu-pairalign}
\bibfield{author}{\bibinfo{person}{Jacek Blazewicz}, \bibinfo{person}{Wojciech Frohmberg}, \bibinfo{person}{Michal Kierzynka}, \bibinfo{person}{Erwin Pesch}, {and} \bibinfo{person}{Pawel Wojciechowski}.} \bibinfo{year}{2011}\natexlab{}.
\newblock \showarticletitle{Protein alignment algorithms with an efficient backtracking routine on multiple {GPU}s}.
\newblock \bibinfo{journal}{\emph{BMC bioinformatics}} \bibinfo{volume}{12}, \bibinfo{number}{1} (\bibinfo{year}{2011}), \bibinfo{pages}{1--17}.
\newblock


\bibitem[Bohra et~al\mbox{.}(2020)]%
        {bohra2020genomic}
\bibfield{author}{\bibinfo{person}{Abhishek Bohra}, \bibinfo{person}{Uday Chand~Jha}, \bibinfo{person}{Ian~D Godwin}, {and} \bibinfo{person}{Rajeev Kumar~Varshney}.} \bibinfo{year}{2020}\natexlab{}.
\newblock \showarticletitle{{Genomic interventions for sustainable agriculture}}.
\newblock \bibinfo{journal}{\emph{Plant Biotechnology Journal}} \bibinfo{volume}{18}, \bibinfo{number}{12} (\bibinfo{year}{2020}), \bibinfo{pages}{2388--2405}.
\newblock


\bibitem[Chatterjee et~al\mbox{.}(2011)]%
        {chatterjee2013dynamic}
\bibfield{author}{\bibinfo{person}{Sanjay Chatterjee}, \bibinfo{person}{Max Grossman}, \bibinfo{person}{Alina Sb{\^\i}rlea}, {and} \bibinfo{person}{Vivek Sarkar}.} \bibinfo{year}{2011}\natexlab{}.
\newblock \showarticletitle{Dynamic task parallelism with a GPU work-stealing runtime system}. In \bibinfo{booktitle}{\emph{LCPC Workshop}}. \bibinfo{publisher}{Springer}, \bibinfo{address}{Fort Collins, CO, USA}, \bibinfo{pages}{203--217}.
\newblock


\bibitem[Choquette(2023)]%
        {hopper}
\bibfield{author}{\bibinfo{person}{Jack Choquette}.} \bibinfo{year}{2023}\natexlab{}.
\newblock \showarticletitle{{NVIDIA Hopper H100 GPU: Scaling Performance}}.
\newblock \bibinfo{journal}{\emph{IEEE Micro}}  \bibinfo{volume}{43} (\bibinfo{year}{2023}), \bibinfo{pages}{9--17}.
\newblock


\bibitem[de~Oliveira~Sandes et~al\mbox{.}(2016)]%
        {cudalign4}
\bibfield{author}{\bibinfo{person}{Edans~Flavius de Oliveira~Sandes}, \bibinfo{person}{Guillermo Miranda}, \bibinfo{person}{Xavier Martorell}, \bibinfo{person}{Eduard Ayguade}, \bibinfo{person}{George Teodoro}, {and} \bibinfo{person}{Alba Cristina~Magalhaes Melo}.} \bibinfo{year}{2016}\natexlab{}.
\newblock \showarticletitle{{CUDAlign} 4.0: Incremental speculative traceback for exact chromosome-wide alignment in {GPU} clusters}.
\newblock \bibinfo{journal}{\emph{IEEE Transactions on Parallel and Distributed Systems}} \bibinfo{volume}{27}, \bibinfo{number}{10} (\bibinfo{year}{2016}), \bibinfo{pages}{2838--2850}.
\newblock


\bibitem[Edans et~al\mbox{.}(2014)]%
        {cudalign3}
\bibfield{author}{\bibinfo{person}{F~de~O Edans}, \bibinfo{person}{Guillermo Miranda}, \bibinfo{person}{Alba~CMA de Melo}, \bibinfo{person}{Xavier Martorell}, {and} \bibinfo{person}{Eduard Ayguad{\'e}}.} \bibinfo{year}{2014}\natexlab{}.
\newblock \showarticletitle{{CUDAlign} 3.0: Parallel biological sequence comparison in large {GPU} clusters}. In \bibinfo{booktitle}{\emph{CCGrid}}. \bibinfo{publisher}{IEEE}, \bibinfo{address}{Chicago, IL, USA}, \bibinfo{pages}{160--169}.
\newblock


\bibitem[Feng et~al\mbox{.}(2019)]%
        {manymap}
\bibfield{author}{\bibinfo{person}{Zonghao Feng}, \bibinfo{person}{Shuang Qiu}, \bibinfo{person}{Lipeng Wang}, {and} \bibinfo{person}{Qiong Luo}.} \bibinfo{year}{2019}\natexlab{}.
\newblock \showarticletitle{Accelerating long read alignment on three processors}. In \bibinfo{booktitle}{\emph{ICPP}}. \bibinfo{publisher}{ACM}, \bibinfo{address}{Kyoto, Japan}, \bibinfo{pages}{1--10}.
\newblock


\bibitem[Fujiki et~al\mbox{.}(2020)]%
        {fujiki2020seedex}
\bibfield{author}{\bibinfo{person}{Daichi Fujiki}, \bibinfo{person}{Shunhao Wu}, \bibinfo{person}{Nathan Ozog}, \bibinfo{person}{Kush Goliya}, \bibinfo{person}{David Blaauw}, \bibinfo{person}{Satish Narayanasamy}, {and} \bibinfo{person}{Reetuparna Das}.} \bibinfo{year}{2020}\natexlab{}.
\newblock \showarticletitle{{SeedEx}: A Genome Sequencing Accelerator for Optimal Alignments in Subminimal Space}. In \bibinfo{booktitle}{\emph{MICRO}}. \bibinfo{publisher}{IEEE}, \bibinfo{address}{Athens, Greece}, \bibinfo{pages}{937--950}.
\newblock


\bibitem[Gamaarachchi et~al\mbox{.}(2020)]%
        {f5c}
\bibfield{author}{\bibinfo{person}{Hasindu Gamaarachchi}, \bibinfo{person}{Chun~Wai Lam}, \bibinfo{person}{Gihan Jayatilaka}, \bibinfo{person}{Hiruna Samarakoon}, \bibinfo{person}{Jared~T Simpson}, \bibinfo{person}{Martin~A Smith}, {and} \bibinfo{person}{Sri Parameswaran}.} \bibinfo{year}{2020}\natexlab{}.
\newblock \showarticletitle{GPU accelerated adaptive banded event alignment for rapid comparative nanopore signal analysis}.
\newblock \bibinfo{journal}{\emph{BMC bioinformatics}}  \bibinfo{volume}{21} (\bibinfo{year}{2020}), \bibinfo{pages}{1--13}.
\newblock


\bibitem[Giani et~al\mbox{.}(2020)]%
        {giani2020long}
\bibfield{author}{\bibinfo{person}{Alice~Maria Giani}, \bibinfo{person}{Guido~Roberto Gallo}, \bibinfo{person}{Luca Gianfranceschi}, {and} \bibinfo{person}{Giulio Formenti}.} \bibinfo{year}{2020}\natexlab{}.
\newblock \showarticletitle{{Long walk to genomics: History and current approaches to genome sequencing and assembly}}.
\newblock \bibinfo{journal}{\emph{Computational and Structural Biotechnology Journal}}  \bibinfo{volume}{18} (\bibinfo{year}{2020}), \bibinfo{pages}{9--19}.
\newblock


\bibitem[Gupta et~al\mbox{.}(2012)]%
        {gupta2012study}
\bibfield{author}{\bibinfo{person}{Kshitij Gupta}, \bibinfo{person}{Jeff~A Stuart}, {and} \bibinfo{person}{John~D Owens}.} \bibinfo{year}{2012}\natexlab{}.
\newblock \showarticletitle{A study of persistent threads style GPU programming for GPGPU workloads}. In \bibinfo{booktitle}{\emph{InPar}}. \bibinfo{publisher}{IEEE}, \bibinfo{address}{San Jose, CA, USA}, \bibinfo{pages}{1--14}.
\newblock


\bibitem[Hong et~al\mbox{.}(2011)]%
        {vwc}
\bibfield{author}{\bibinfo{person}{Sungpack Hong}, \bibinfo{person}{Sang~Kyun Kim}, \bibinfo{person}{Tayo Oguntebi}, {and} \bibinfo{person}{Kunle Olukotun}.} \bibinfo{year}{2011}\natexlab{}.
\newblock \showarticletitle{{Accelerating CUDA graph algorithms at maximum warp}}. In \bibinfo{booktitle}{\emph{PPoPP}}. \bibinfo{publisher}{ACM}, \bibinfo{address}{San Antonio, TX, USA}, \bibinfo{pages}{267--276}.
\newblock


\bibitem[Kalikar et~al\mbox{.}(2022)]%
        {minimap2avx}
\bibfield{author}{\bibinfo{person}{Saurabh Kalikar}, \bibinfo{person}{Chirag Jain}, \bibinfo{person}{Md Vasimuddin}, {and} \bibinfo{person}{Sanchit Misra}.} \bibinfo{year}{2022}\natexlab{}.
\newblock \showarticletitle{Accelerating minimap2 for long-read sequencing applications on modern CPUs}.
\newblock \bibinfo{journal}{\emph{Nature Computational Science}} \bibinfo{volume}{2}, \bibinfo{number}{2} (\bibinfo{year}{2022}), \bibinfo{pages}{78--83}.
\newblock


\bibitem[Khorasani et~al\mbox{.}(2015)]%
        {khorasani2015efficient}
\bibfield{author}{\bibinfo{person}{Farzad Khorasani}, \bibinfo{person}{Rajiv Gupta}, {and} \bibinfo{person}{Laxmi~N Bhuyan}.} \bibinfo{year}{2015}\natexlab{}.
\newblock \showarticletitle{Efficient warp execution in presence of divergence with collaborative context collection}. In \bibinfo{booktitle}{\emph{MICRO}}. \bibinfo{publisher}{ACM}, \bibinfo{address}{Waikiki, HI, USA}, \bibinfo{pages}{204--215}.
\newblock


\bibitem[Khorasani et~al\mbox{.}(2016)]%
        {khorasani2016eliminating}
\bibfield{author}{\bibinfo{person}{Farzad Khorasani}, \bibinfo{person}{Bryan Rowe}, \bibinfo{person}{Rajiv Gupta}, {and} \bibinfo{person}{Laxmi~N Bhuyan}.} \bibinfo{year}{2016}\natexlab{}.
\newblock \showarticletitle{Eliminating intra-warp load imbalance in irregular nested patterns via collaborative task engagement}. In \bibinfo{booktitle}{\emph{IPDPS}}. \bibinfo{publisher}{IEEE}, \bibinfo{address}{Chicago, IL, USA}, \bibinfo{pages}{524--533}.
\newblock


\bibitem[Korpar and {\v{S}}iki{\'c}(2013)]%
        {korpar2013sw}
\bibfield{author}{\bibinfo{person}{Matija Korpar} {and} \bibinfo{person}{Mile {\v{S}}iki{\'c}}.} \bibinfo{year}{2013}\natexlab{}.
\newblock \showarticletitle{{SW\#}--{GPU}-enabled exact alignments on genome scale}.
\newblock \bibinfo{journal}{\emph{Bioinformatics}} \bibinfo{volume}{29}, \bibinfo{number}{19} (\bibinfo{year}{2013}), \bibinfo{pages}{2494--2495}.
\newblock


\bibitem[Li(2013)]%
        {bwamem}
\bibfield{author}{\bibinfo{person}{Heng Li}.} \bibinfo{year}{2013}\natexlab{}.
\newblock \bibinfo{title}{Aligning sequence reads, clone sequences and assembly contigs with BWA-MEM}.
\newblock
\newblock
\showeprint[arxiv]{1303.3997}~[q-bio.GN]


\bibitem[Li(2018)]%
        {minimap2}
\bibfield{author}{\bibinfo{person}{Heng Li}.} \bibinfo{year}{2018}\natexlab{}.
\newblock \showarticletitle{Minimap2: Pairwise alignment for nucleotide sequences}.
\newblock \bibinfo{journal}{\emph{Bioinformatics}} \bibinfo{volume}{34}, \bibinfo{number}{18} (\bibinfo{year}{2018}), \bibinfo{pages}{3094--3100}.
\newblock


\bibitem[Liu et~al\mbox{.}(2012b)]%
        {soap3}
\bibfield{author}{\bibinfo{person}{Chi-Man Liu}, \bibinfo{person}{Thomas Wong}, \bibinfo{person}{Edward Wu}, \bibinfo{person}{Ruibang Luo}, \bibinfo{person}{Siu-Ming Yiu}, \bibinfo{person}{Yingrui Li}, \bibinfo{person}{Bingqiang Wang}, \bibinfo{person}{Chang Yu}, \bibinfo{person}{Xiaowen Chu}, \bibinfo{person}{Kaiyong Zhao}, {et~al\mbox{.}}} \bibinfo{year}{2012}\natexlab{b}.
\newblock \showarticletitle{{SOAP}3: Ultra-fast {GPU}-based parallel alignment tool for short reads}.
\newblock \bibinfo{journal}{\emph{Bioinformatics}} \bibinfo{volume}{28}, \bibinfo{number}{6} (\bibinfo{year}{2012}), \bibinfo{pages}{878--879}.
\newblock


\bibitem[Liu et~al\mbox{.}(2006b)]%
        {liu2006bio}
\bibfield{author}{\bibinfo{person}{Weiguo Liu}, \bibinfo{person}{Bertil Schmidt}, \bibinfo{person}{Gerrit Voss}, \bibinfo{person}{Andre Schroder}, {and} \bibinfo{person}{Wolfgang Muller-Wittig}.} \bibinfo{year}{2006}\natexlab{b}.
\newblock \showarticletitle{{Bio-sequence database scanning on a {GPU}}}. In \bibinfo{booktitle}{\emph{IPDPS}}. \bibinfo{publisher}{IEEE}, \bibinfo{address}{Rhodes Island, Greece}, \bibinfo{pages}{1--8}.
\newblock


\bibitem[Liu et~al\mbox{.}(2006a)]%
        {liu2006gpu}
\bibfield{author}{\bibinfo{person}{Yang Liu}, \bibinfo{person}{Wayne Huang}, \bibinfo{person}{John Johnson}, {and} \bibinfo{person}{Sheila Vaidya}.} \bibinfo{year}{2006}\natexlab{a}.
\newblock \showarticletitle{GPU accelerated smith-waterman}. In \bibinfo{booktitle}{\emph{ICCS}}. \bibinfo{publisher}{Springer}, \bibinfo{address}{Reading, UK}, \bibinfo{pages}{188--195}.
\newblock


\bibitem[Liu et~al\mbox{.}(2009)]%
        {cudasw1}
\bibfield{author}{\bibinfo{person}{Yongchao Liu}, \bibinfo{person}{Douglas~L Maskell}, {and} \bibinfo{person}{Bertil Schmidt}.} \bibinfo{year}{2009}\natexlab{}.
\newblock \showarticletitle{{CUDASW}++: Optimizing {Smith-Waterman} sequence database searches for {CUDA}-enabled graphics processing units}.
\newblock \bibinfo{journal}{\emph{BMC research notes}} \bibinfo{volume}{2}, \bibinfo{number}{1} (\bibinfo{year}{2009}), \bibinfo{pages}{1--10}.
\newblock


\bibitem[Liu et~al\mbox{.}(2013a)]%
        {cushaw2}
\bibfield{author}{\bibinfo{person}{Yongchao Liu}, \bibinfo{person}{Bernt Popp}, {and} \bibinfo{person}{Bertil Schmidt}.} \bibinfo{year}{2013}\natexlab{a}.
\newblock \bibinfo{title}{High-speed and accurate color-space short-read alignment with CUSHAW2}.
\newblock
\newblock
\showeprint[arxiv]{1304.4766}~[q-bio.GN]


\bibitem[Liu et~al\mbox{.}(2014)]%
        {cushaw3}
\bibfield{author}{\bibinfo{person}{Yongchao Liu}, \bibinfo{person}{Bernt Popp}, {and} \bibinfo{person}{Bertil Schmidt}.} \bibinfo{year}{2014}\natexlab{}.
\newblock \showarticletitle{{CUSHAW3}: Sensitive and accurate base-space and color-space short-read alignment with hybrid seeding}.
\newblock \bibinfo{journal}{\emph{PloS one}} \bibinfo{volume}{9}, \bibinfo{number}{1} (\bibinfo{year}{2014}), \bibinfo{pages}{e86869}.
\newblock


\bibitem[Liu and Schmidt(2015)]%
        {gswabe}
\bibfield{author}{\bibinfo{person}{Yongchao Liu} {and} \bibinfo{person}{Bertil Schmidt}.} \bibinfo{year}{2015}\natexlab{}.
\newblock \showarticletitle{{GSWABE}: Faster {GPU}-accelerated sequence alignment with optimal alignment retrieval for short {DNA} sequences}.
\newblock \bibinfo{journal}{\emph{Concurrency and Computation: Practice and Experience}} \bibinfo{volume}{27}, \bibinfo{number}{4} (\bibinfo{year}{2015}), \bibinfo{pages}{958--972}.
\newblock


\bibitem[Liu et~al\mbox{.}(2010)]%
        {cudasw2}
\bibfield{author}{\bibinfo{person}{Yongchao Liu}, \bibinfo{person}{Bertil Schmidt}, {and} \bibinfo{person}{Douglas~L Maskell}.} \bibinfo{year}{2010}\natexlab{}.
\newblock \showarticletitle{{CUDASW}++ 2.0: Enhanced Smith-Waterman protein database search on {CUDA}-enabled {GPU}s based on {SIMT} and virtualized {SIMD} abstractions}.
\newblock \bibinfo{journal}{\emph{BMC research notes}} \bibinfo{volume}{3}, \bibinfo{number}{1} (\bibinfo{year}{2010}), \bibinfo{pages}{1--12}.
\newblock


\bibitem[Liu et~al\mbox{.}(2012a)]%
        {cushaw1}
\bibfield{author}{\bibinfo{person}{Yongchao Liu}, \bibinfo{person}{Bertil Schmidt}, {and} \bibinfo{person}{Douglas~L Maskell}.} \bibinfo{year}{2012}\natexlab{a}.
\newblock \showarticletitle{{CUSHAW}: A {CUDA} compatible short read aligner to large genomes based on the {Burrows--Wheeler} transform}.
\newblock \bibinfo{journal}{\emph{Bioinformatics}} \bibinfo{volume}{28}, \bibinfo{number}{14} (\bibinfo{year}{2012}), \bibinfo{pages}{1830--1837}.
\newblock


\bibitem[Liu et~al\mbox{.}(2013b)]%
        {cudasw3}
\bibfield{author}{\bibinfo{person}{Yongchao Liu}, \bibinfo{person}{Adrianto Wirawan}, {and} \bibinfo{person}{Bertil Schmidt}.} \bibinfo{year}{2013}\natexlab{b}.
\newblock \showarticletitle{{CUDASW}++ 3.0: Accelerating {Smith-Waterman} protein database search by coupling {CPU} and {GPU} {SIMD} instructions}.
\newblock \bibinfo{journal}{\emph{BMC bioinformatics}} \bibinfo{volume}{14}, \bibinfo{number}{1} (\bibinfo{year}{2013}), \bibinfo{pages}{1--10}.
\newblock


\bibitem[Mantere et~al\mbox{.}(2019)]%
        {mantere2019long}
\bibfield{author}{\bibinfo{person}{Tuomo Mantere}, \bibinfo{person}{Simone Kersten}, {and} \bibinfo{person}{Alexander Hoischen}.} \bibinfo{year}{2019}\natexlab{}.
\newblock \showarticletitle{{Long-read sequencing emerging in medical genetics}}.
\newblock \bibinfo{journal}{\emph{Frontiers in genetics}}  \bibinfo{volume}{10} (\bibinfo{year}{2019}), \bibinfo{pages}{426}.
\newblock


\bibitem[NCBI(1982)]%
        {genbank}
\bibfield{author}{\bibinfo{person}{NCBI}.} \bibinfo{year}{1982}\natexlab{}.
\newblock \bibinfo{title}{{GenBank}}.
\newblock
\newblock
\newblock
\shownote{\url{https://www.ncbi.nlm.nih.gov/genbank/}, visited 2024-01-15}.


\bibitem[Needleman and Wunsch(1970)]%
        {nw}
\bibfield{author}{\bibinfo{person}{Saul~B Needleman} {and} \bibinfo{person}{Christian~D Wunsch}.} \bibinfo{year}{1970}\natexlab{}.
\newblock \showarticletitle{{A general method applicable to the search for similarities in the amino acid sequence of two proteins}}.
\newblock \bibinfo{journal}{\emph{Journal of molecular biology}} \bibinfo{volume}{48}, \bibinfo{number}{3} (\bibinfo{year}{1970}), \bibinfo{pages}{443--453}.
\newblock


\bibitem[NIST(2012)]%
        {genome-in-a-bottle}
\bibfield{author}{\bibinfo{person}{NIST}.} \bibinfo{year}{2012}\natexlab{}.
\newblock \bibinfo{title}{{Genome in a Bottle}}.
\newblock
\newblock
\newblock
\shownote{\url{https://www.nist.gov/programs-projects/genome-bottle}, visited 2024-01-16}.


\bibitem[NVIDIA(2022a)]%
        {dpxsw}
\bibfield{author}{\bibinfo{person}{NVIDIA}.} \bibinfo{year}{2022}\natexlab{a}.
\newblock \bibinfo{title}{{Boosting Dynamic Programming Performance Using NVIDIA Hopper GPU DPX Instructions}}.
\newblock
\newblock
\newblock
\shownote{\url{https://developer.nvidia.com/blog/boosting-dynamic-programming-performance-using-nvidia-hopper-gpu-dpx-instructions/}, visited 2024-01-16}.


\bibitem[NVIDIA(2022b)]%
        {dpx}
\bibfield{author}{\bibinfo{person}{NVIDIA}.} \bibinfo{year}{2022}\natexlab{b}.
\newblock \bibinfo{title}{{NVIDIA Hopper GPU Architecture Accelerates Dynamic Programming Up to 40x Using New DPX Instructions}}.
\newblock
\newblock
\newblock
\shownote{\url{https://blogs.nvidia.com/blog/2022/03/22/nvidia-hopper-accelerates-dynamic-programming-using-dpx-instructions/}, visited 2024-01-15}.


\bibitem[PacBio(2023)]%
        {clr}
\bibfield{author}{\bibinfo{person}{PacBio}.} \bibinfo{year}{2023}\natexlab{}.
\newblock \bibinfo{title}{{PacBio - Sequence with Confidence}}.
\newblock
\newblock
\newblock
\shownote{\url{https://www.pacb.com/}, visited 2023-04-21}.


\bibitem[Park(2024)]%
        {agatha}
\bibfield{author}{\bibinfo{person}{Seongyeon Park}.} \bibinfo{year}{2024}\natexlab{}.
\newblock \bibinfo{booktitle}{\emph{{readwrite112/AGAThA: AGAThA: Fast and Efficient GPU Acceleration of Guided Sequence Alignment for Long Read Mapping}}}.
\newblock AISys.
\newblock
\urldef\tempurl%
\url{https://doi.org/10.5281/zenodo.10462237}
\showDOI{\tempurl}


\bibitem[Park et~al\mbox{.}(2022)]%
        {saloba}
\bibfield{author}{\bibinfo{person}{Seongyeon Park}, \bibinfo{person}{Hajin Kim}, \bibinfo{person}{Tanveer Ahmad}, \bibinfo{person}{Nauman Ahmed}, \bibinfo{person}{Zaid Al-Ars}, \bibinfo{person}{H~Peter Hofstee}, \bibinfo{person}{Youngsok Kim}, {and} \bibinfo{person}{Jinho Lee}.} \bibinfo{year}{2022}\natexlab{}.
\newblock \showarticletitle{SALoBa: Maximizing Data Locality and Workload Balance for Fast Sequence Alignment on GPUs}. In \bibinfo{booktitle}{\emph{IPDPS}}. \bibinfo{publisher}{IEEE}, \bibinfo{address}{Lyon, France}, \bibinfo{pages}{728--738}.
\newblock


\bibitem[Parkhill and Wren(2011)]%
        {parkhill2011bacterial}
\bibfield{author}{\bibinfo{person}{Julian Parkhill} {and} \bibinfo{person}{Brendan~W Wren}.} \bibinfo{year}{2011}\natexlab{}.
\newblock \showarticletitle{{Bacterial epidemiology and biology-lessons from genome sequencing}}.
\newblock \bibinfo{journal}{\emph{Genome biology}}  \bibinfo{volume}{12} (\bibinfo{year}{2011}), \bibinfo{pages}{1--7}.
\newblock


\bibitem[RefSeq(2019)]%
        {refsequence}
\bibfield{author}{\bibinfo{person}{NCBI RefSeq}.} \bibinfo{year}{2019}\natexlab{}.
\newblock \bibinfo{title}{{GRCh38.p13}}.
\newblock
\newblock
\newblock
\shownote{\url{https://www.ncbi.nlm.nih.gov/data-hub/genome/GCF_000001405.39/}, visited 2024-01-16}.


\bibitem[Sadasivan et~al\mbox{.}(2023)]%
        {sadasivan2023accelerating}
\bibfield{author}{\bibinfo{person}{Harisankar Sadasivan}, \bibinfo{person}{Milos Maric}, \bibinfo{person}{Eric Dawson}, \bibinfo{person}{Vishanth Iyer}, \bibinfo{person}{Johnny Israeli}, {and} \bibinfo{person}{Satish Narayanasamy}.} \bibinfo{year}{2023}\natexlab{}.
\newblock \showarticletitle{Accelerating Minimap2 for accurate long read alignment on GPUs}.
\newblock \bibinfo{journal}{\emph{Journal of biotechnology and biomedicine}} \bibinfo{volume}{6}, \bibinfo{number}{1} (\bibinfo{year}{2023}), \bibinfo{pages}{13--23}.
\newblock


\bibitem[Sandes and de~Melo(2011)]%
        {cudalign2}
\bibfield{author}{\bibinfo{person}{Edans Flavius de~O Sandes} {and} \bibinfo{person}{Alba Cristina~MA de Melo}.} \bibinfo{year}{2011}\natexlab{}.
\newblock \showarticletitle{{Smith-Waterman} alignment of huge sequences with {GPU} in linear space}. In \bibinfo{booktitle}{\emph{IPDPS}}. \bibinfo{publisher}{IEEE}, \bibinfo{address}{Anchorage, AK, USA}, \bibinfo{pages}{1199--1211}.
\newblock


\bibitem[Sandes and de~Melo(2010)]%
        {cudalign1}
\bibfield{author}{\bibinfo{person}{Edans Flavius~O Sandes} {and} \bibinfo{person}{Alba Cristina~MA de Melo}.} \bibinfo{year}{2010}\natexlab{}.
\newblock \showarticletitle{{CUDAlign}: Using {GPU} to accelerate the comparison of megabase genomic sequences}. In \bibinfo{booktitle}{\emph{PPoPP}}. \bibinfo{publisher}{ACM}, \bibinfo{address}{New York, NY, USA}, \bibinfo{pages}{137--146}.
\newblock


\bibitem[Smith et~al\mbox{.}(1981)]%
        {sw}
\bibfield{author}{\bibinfo{person}{Temple~F Smith}, \bibinfo{person}{Michael~S Waterman}, {et~al\mbox{.}}} \bibinfo{year}{1981}\natexlab{}.
\newblock \showarticletitle{{Identification of common molecular subsequences}}.
\newblock \bibinfo{journal}{\emph{Journal of molecular biology}} \bibinfo{volume}{147}, \bibinfo{number}{1} (\bibinfo{year}{1981}), \bibinfo{pages}{195--197}.
\newblock


\bibitem[Steinberger et~al\mbox{.}(2012)]%
        {steinberger2012softshell}
\bibfield{author}{\bibinfo{person}{Markus Steinberger}, \bibinfo{person}{Bernhard Kainz}, \bibinfo{person}{Bernhard Kerbl}, \bibinfo{person}{Stefan Hauswiesner}, \bibinfo{person}{Michael Kenzel}, {and} \bibinfo{person}{Dieter Schmalstieg}.} \bibinfo{year}{2012}\natexlab{}.
\newblock \showarticletitle{Softshell: dynamic scheduling on gpus}.
\newblock \bibinfo{journal}{\emph{ACM Transactions on Graphics}} \bibinfo{volume}{31}, \bibinfo{number}{6} (\bibinfo{year}{2012}), \bibinfo{pages}{1--11}.
\newblock


\bibitem[Steinberger et~al\mbox{.}(2014)]%
        {steinberger2014whippletree}
\bibfield{author}{\bibinfo{person}{Markus Steinberger}, \bibinfo{person}{Michael Kenzel}, \bibinfo{person}{Pedro Boechat}, \bibinfo{person}{Bernhard Kerbl}, \bibinfo{person}{Mark Dokter}, {and} \bibinfo{person}{Dieter Schmalstieg}.} \bibinfo{year}{2014}\natexlab{}.
\newblock \showarticletitle{Whippletree: Task-based scheduling of dynamic workloads on the GPU}.
\newblock \bibinfo{journal}{\emph{ACM Transactions on Graphics}} \bibinfo{volume}{33}, \bibinfo{number}{6} (\bibinfo{year}{2014}), \bibinfo{pages}{1--11}.
\newblock


\bibitem[Suzuki(2020)]%
        {suzuki2020advent}
\bibfield{author}{\bibinfo{person}{Yutaka Suzuki}.} \bibinfo{year}{2020}\natexlab{}.
\newblock \showarticletitle{Advent of a new sequencing era: long-read and on-site sequencing}.
\newblock \bibinfo{journal}{\emph{Journal of Human Genetics}} \bibinfo{volume}{65}, \bibinfo{number}{1} (\bibinfo{year}{2020}), \bibinfo{pages}{1--1}.
\newblock


\bibitem[Technologies(2018)]%
        {ont}
\bibfield{author}{\bibinfo{person}{Oxford~Nanopore Technologies}.} \bibinfo{year}{2018}\natexlab{}.
\newblock \bibinfo{title}{{PromethION}}.
\newblock
\newblock
\newblock
\shownote{\url{https://nanoporetech.com/products/promethion}, visited 2024-01-16}.


\bibitem[Tzeng et~al\mbox{.}(2010)]%
        {tzeng2010task}
\bibfield{author}{\bibinfo{person}{Stanley Tzeng}, \bibinfo{person}{Anjul Patney}, {and} \bibinfo{person}{John~D Owens}.} \bibinfo{year}{2010}\natexlab{}.
\newblock \showarticletitle{Task management for irregular-parallel workloads on the GPU}. In \bibinfo{booktitle}{\emph{HPG}}. \bibinfo{publisher}{ACM}, \bibinfo{address}{Saarbrucken, Germany}, \bibinfo{pages}{29--37}.
\newblock


\bibitem[Wei et~al\mbox{.}(2020)]%
        {smsmap}
\bibfield{author}{\bibinfo{person}{Ze-Gang Wei}, \bibinfo{person}{Shao-Wu Zhang}, {and} \bibinfo{person}{Fei Liu}.} \bibinfo{year}{2020}\natexlab{}.
\newblock \showarticletitle{smsMap: mapping single molecule sequencing reads by locating the alignment starting positions}.
\newblock \bibinfo{journal}{\emph{BMC bioinformatics}} \bibinfo{volume}{21}, \bibinfo{number}{1} (\bibinfo{year}{2020}), \bibinfo{pages}{1--15}.
\newblock


\bibitem[Wenger et~al\mbox{.}(2019)]%
        {hifi}
\bibfield{author}{\bibinfo{person}{Aaron~M Wenger}, \bibinfo{person}{Paul Peluso}, \bibinfo{person}{William~J Rowell}, \bibinfo{person}{Pi-Chuan Chang}, \bibinfo{person}{Richard~J Hall}, \bibinfo{person}{Gregory~T Concepcion}, \bibinfo{person}{Jana Ebler}, \bibinfo{person}{Arkarachai Fungtammasan}, \bibinfo{person}{Alexey Kolesnikov}, \bibinfo{person}{Nathan~D Olson}, {et~al\mbox{.}}} \bibinfo{year}{2019}\natexlab{}.
\newblock \showarticletitle{{Accurate circular consensus long-read sequencing improves variant detection and assembly of a human genome}}.
\newblock \bibinfo{journal}{\emph{Nature biotechnology}} \bibinfo{volume}{37}, \bibinfo{number}{10} (\bibinfo{year}{2019}), \bibinfo{pages}{1155--1162}.
\newblock


\bibitem[Wu et~al\mbox{.}(2015)]%
        {wu2015enabling}
\bibfield{author}{\bibinfo{person}{Bo Wu}, \bibinfo{person}{Guoyang Chen}, \bibinfo{person}{Dong Li}, \bibinfo{person}{Xipeng Shen}, {and} \bibinfo{person}{Jeffrey Vetter}.} \bibinfo{year}{2015}\natexlab{}.
\newblock \showarticletitle{Enabling and exploiting flexible task assignment on GPU through SM-centric program transformations}. In \bibinfo{booktitle}{\emph{ICS}}. \bibinfo{publisher}{ACM}, \bibinfo{address}{Newport Beach, CA, USA}, \bibinfo{pages}{119--130}.
\newblock


\bibitem[Zeni et~al\mbox{.}(2020)]%
        {zeni2020logan}
\bibfield{author}{\bibinfo{person}{Alberto Zeni}, \bibinfo{person}{Giulia Guidi}, \bibinfo{person}{Marquita Ellis}, \bibinfo{person}{Nan Ding}, \bibinfo{person}{Marco~D Santambrogio}, \bibinfo{person}{Steven Hofmeyr}, \bibinfo{person}{Ayd{\i}n Bulu{\c{c}}}, \bibinfo{person}{Leonid Oliker}, {and} \bibinfo{person}{Katherine Yelick}.} \bibinfo{year}{2020}\natexlab{}.
\newblock \showarticletitle{{LOGAN}: High-performance {GPU}-based x-drop long-read alignment}. In \bibinfo{booktitle}{\emph{IPDPS}}. \bibinfo{publisher}{IEEE}, \bibinfo{address}{New Orleans, LA, USA}, \bibinfo{pages}{462--471}.
\newblock


\bibitem[Zhang et~al\mbox{.}(2010)]%
        {zhang2010streamlining}
\bibfield{author}{\bibinfo{person}{Eddy~Z Zhang}, \bibinfo{person}{Yunlian Jiang}, \bibinfo{person}{Ziyu Guo}, {and} \bibinfo{person}{Xipeng Shen}.} \bibinfo{year}{2010}\natexlab{}.
\newblock \showarticletitle{Streamlining GPU applications on the fly: thread divergence elimination through runtime thread-data remapping}. In \bibinfo{booktitle}{\emph{ICS}}. \bibinfo{publisher}{ACM}, \bibinfo{address}{Tsukuba, Ibaraki, Japan}, \bibinfo{pages}{115--126}.
\newblock


\bibitem[Zheng et~al\mbox{.}(2017)]%
        {zheng2017versapipe}
\bibfield{author}{\bibinfo{person}{Zhen Zheng}, \bibinfo{person}{Chanyoung Oh}, \bibinfo{person}{Jidong Zhai}, \bibinfo{person}{Xipeng Shen}, \bibinfo{person}{Youngmin Yi}, {and} \bibinfo{person}{Wenguang Chen}.} \bibinfo{year}{2017}\natexlab{}.
\newblock \showarticletitle{Versapipe: a versatile programming framework for pipelined computing on GPU}. In \bibinfo{booktitle}{\emph{MICRO}}. \bibinfo{publisher}{ACM}, \bibinfo{address}{Cambridge, MA, USA}, \bibinfo{pages}{587--599}.
\newblock


\end{thebibliography}

\newpage

\appendix
\section{Appendix}
\label{sec:appendix}

We provide \thiswork's source code, additional code for setup and execution, and a sample dataset as an example for the input format. 
For the most recent version of \thiswork's description, please refer to the up-to-date artifact link in GitHub.

\subsection{Artifact Summary}

\begin{itemize}
    \item \textbf{Dataset}: A sample of HiFi HG005 dataset. 
    \item \textbf{Runtime environment}: Ubuntu 20.04 or higher.
    \item \textbf{Hardware}: Multicore x86\_64 CPU with one or more NVIDIA GPUs. 
    \item \textbf{Metrics}: Execution time.
    \item \textbf{Code license}: Apache-2.0 license.
    \item \textbf{Archived DOI}:\\\url{https://doi.org/10.5281/zenodo.10462237}
    \item \textbf{Up-to-date artifact}: \\\url{https://github.com/readwrite112/AGAThA}
\end{itemize}

\subsection{Description}
\subsubsection{How to access.}
Please access the artifact via the archived DOI~\cite{agatha} or the up-to-date artifact link.
\subsubsection{Hardware dependencies.}
Requires multicore x86\_64 CPU with at least a single NVIDIA GPU. The artifact was tested on NVIDIA RTX A6000 and NVIDIA GeForce RTX 4090, both with NVIDIA compute capability 86 (sm86).  


\subsubsection{Artifact structure}
The artifact's structure can be represented as the following:
\begin{verbatim}
    |-- AGAThA
    |   |-- src
    |   |   `-- kernels
    |   `-- test_prog
    |-- dataset
    |-- docker
    |-- misc
    `-- output
\end{verbatim}
\begin{itemize}
    \item \texttt{\thiswork/src}: source code for \thiswork.
    \item \texttt{\thiswork/test\_prog}: test program to use \thiswork's kernels for sequence alignment.
    \item \texttt{dataset}: includes sample input dataset. 
    \item \texttt{docker}: scripts and Dockerfile for launching docker.
    \item \texttt{misc}: miscellaneous code for outputting \thiswork's kernel execution time.
    \item \texttt{output}: directory for outputs such as alignment score and kernel execution time. 
\end{itemize}

\subsubsection{Setup.}
We recommend building and launching a docker image with the following scripts.
\begin{verbatim}
    $ cd docker
    $ bash build.sh
    $ bash launch.sh
\end{verbatim}
\thiswork's source code can be built using the following command lines. 
\begin{verbatim}
    $ cd AGAThA
    $ bash build.sh 
\end{verbatim}

\subsubsection{Datasets.}
A sample of the HiFi HG005 dataset used in this work is provided in \texttt{dataset/}.
Other (custom) datasets can be used as input, as long as there is a reference and query file, and both are formatted as \texttt{.fasta} files. 
The input files should both follow the format below:
\begin{verbatim}
    >>> 1
    ATGCN...
    >>> 2
    TCGGA...
\end{verbatim}
Given the two input files, \thiswork will output the alignment score between a pair of sequences from each file. 
Note that each input file should have an equal number of reference and query strings.
\texttt{.fasta} files can be download from various sources such as GenBank~\cite{genbank} or projects such as `Genome in a Bottle'~\cite{genome-in-a-bottle}. 

\subsubsection{Running \thiswork}
Using \texttt{AGAThA.sh} script, the following options can be used for \thiswork.
\begin{verbatim}
    -a  the match score
    -b  the mismatch penatly
    -q  the gap open penalty
    -r  the gap extension penalty
    -z  the termination threshold
    -w  the band width in the score table
\end{verbatim}
This script stores the alignment scores in
\texttt{output/score.log},
and the total kernel execution millisecond time is stored in \texttt{output/time.json}.

\end{document}